\documentclass[12pt,preprint]{aastex}

\shorttitle{IC\,133 in M33}
\shortauthors{Argon et al.}

\newcommand\as    {\ifmmode {\rlap.}^{''}\! \else ${\rlap.}^{''}\!$\fi}
\newcommand\porm  {\ifmmode {~\pm~} \else {$~\pm~$} \fi}
\newcommand\kms   {\ifmmode {{\rm km s}^{-1}}\else{km s$^{-1}$}\fi}
\newcommand\msun  {\ifmmode {{\rm M}_\odot}\else{${\rm M}_\odot$}\fi}
\newcommand\lsun  {\ifmmode {{\rm L}_\odot}\else{${\rm L}_\odot$}\fi}

\begin{document}



\title{The IC\,133 Water Vapor Maser in the Galaxy M33: \\ 
    A Geometric Distance}


\author{A.~L.~Argon, L.~J.~Greenhill\altaffilmark{1}, J.~M.~Moran, and M.~J.~Reid}
\affil{Harvard-Smithsonian Center for Astrophysics, Cambridge, MA}

\author{K.~M.~Menten}
\affil{Max--Planck--Institut f\"ur Radioastronomie, Bonn, Germany}

\and

\author{M.~Inoue}
\affil{Nobeyama Radio Observatory, Nobeyama, Japan}



\altaffiltext{1}{Current Address: Kavli Institute for Particle Astrophysics 
and Cosmology, Stanford Linear Accelerator Center, Mail Stop 29, 2575 Sand 
Hill Rd, Menlo Park, CA 94025.}


\begin{abstract}
We report on the results of a 14 year long VLBI study of proper motions in 
the IC\,133 H$_2$O maser source in the galaxy M33.  The method of {\it 
Ordered Motion Parallax} was used to model the 3-dimensional structure and 
dynamics of IC\,133 and obtain a distance estimate, 800 $\pm$ 180 kpc.  
Our technique for determining the distance to M33 is independent of 
calibrations common to other distance indicators, such as Cepheid 
Period-Luminosity relations, and therefore provides an important check for 
previous distance determinations.
\end{abstract}






\section{Introduction}
Calibration of the extragalactic distance scale (EDS) is of crucial importance 
to modern (precision) cosmology.  Robust estimates of the local expansion rate 
of the universe, H$_\circ$, bear on calculation of cosmological parameters  
such as flatness, the Cosmological constant, and the equation of state of dark 
energy \citep[e.g.,][]{fre01,spe03}.  However, estimation of the EDS involves 
layers of calibration and a variety of techniques, objects, and effects, 
which makes assessment of systematic error difficult.

The best estimate of H$_\circ$ to date has been provided by the Hubble Space 
Telescope (HST) Key Project \citep[hereafter referred to as the Key 
Project;][]{fre01}.  Key Project researchers first calibrated 
Period-Luminosity (PL) relations for Cepheid variable stars in the Large 
Magellanic Cloud (LMC) and applied them to hundreds of variables observed with 
the HST in 31 relatively nearby ($<$ 30 Mpc) galaxies.  They then used the 
resulting distance estimates to bootstrap calibration of other distance 
indicators (e.g., Tully-Fisher) in the range 60$-$400 Mpc.  H$_\circ$ was 
estimated from the entire sample of galaxies.

Conceptually, the weakest link in determination of the EDS is that it
depends on the distance to a single galaxy, the LMC, which remains
controversial.  So-called short and long distances have been obtained from
analyses of a variety of datasets for distance indicators such as detached
eclipsing binaries and red clump stars.  The uncertainty in the distance to 
the LMC may be as small as 5\% (Key Project), but the spread in distance 
estimates is larger than the formal errors, which suggests an uncertainty a 
few times larger than this 
\citep[e.g.,][]{fc97,fit03,gui98,kan03,sta00,uda00}.  Other problems arise 
from the poorly understood structure of the LMC \citep[and references 
therein]{vdm04} and uncertainty in the effect of metalicity on PL relations 
\citep[and references therein]{gro04}, the latter being important in as much 
as the LMC is relatively metal poor compared to galaxies targeted by the Key 
Project.

Prior to the work reported here, NGC\,4258 was the only galaxy for which a 
distance had been estimated from the analysis of H$_2$O maser source structure 
and dynamics.  For NGC\,4258, proper motions and accelerations of maser 
components were measured and fit to a model of a thin disk with circular 
orbits in Keplerian rotation.  A distance estimate of 7.3 $\pm$ 0.3 $\pm$ 
0.4 Mpc was obtained \citep{her99}.  The two errors are statistical 
and systematic, respectively, with the latter being due in large part to 
a relatively weak constraint on orbital eccentricity.  The maser distance 
estimate is very important, since it provides a check of the supporting 
calibrations in Cepheid and other non-maser distance estimates.  The NGC\,4258 
maser distance was found to agree reasonably well with the Key Project 
distance (7.8 $\pm$ 0.3 $\pm$ 0.5 Mpc), after adoption of PL relations derived 
from Optical Gravitational Lensing Experiment (OGLE) data 
\citep{new01,fre01}.  However, the Cepheid data for NGC\,4258 is rather poor 
(e.g., only 15 comparatively metal rich stars with a limited range of 
periods), and it remains possible that some terms in the Key Project error 
budget are larger than originally believed.

Cepheid and maser distances to NGC\,4258 and other nearby galaxies would 
permit robust quantification of the EDS.  Detectable 
H$_2$O maser emission within Local Group galaxies is comparatively rare; it is 
known in M33, IC\,10, and the Magellanic Clouds.  Because it is relatively 
nearby, M33 is an important target for such study.  The IC\,133 maser in a 
star forming region of M33 is only of moderate strength ($\sim$ 1$-$2 Jy), but 
it exhibits numerous Doppler components spread over $\approx$ 50 km s$^{-1}$, 
which is important for the construction of accurate dynamical models.

M33 has been the target of many previous non-maser distance studies.  Table 
1 gives a representative sampling of recent distance estimates to this Local 
Group galaxy.  Many of the quoted errors appear to be small, $\approx$ 5\%, 
but none of the techniques employed were independent of the assumptions and 
calibrations that plague all primary and secondary distance indicators.  
In this paper we report on the results of a VLBI study of H$_2$O masers in 
IC\,133.  As with the NGC\,4258 maser distance, the method employed here is 
independent of the calibrations implicit in previous optical distance 
determinations. 

Spectral line VLBI observations of H$_2$O maser proper motions have been used 
successfully to find distances to a number of high mass star-forming regions 
in our own Galaxy.  These include the Orion Kleinmann-Low (KL) nebula 
\citep{grm81}, W51 Main \citep{gen81}, W51 North \citep{sld81}, Sgr B2 North 
\citep{rsm88}, and W49\,N \citep{gmr92}.  In some of the Galactic star-forming 
regions maser motions appeared random, so a simple statistical comparison of 
the spread in radial velocities to the spread in transverse velocities 
(derived from proper motions) was made to obtain a distance.  In other sources 
maser motions appeared ordered and a dynamical model was fit to the observed 
proper motions and radial velocity distribution, giving distance as a 
parameter of the fit.  The techniques are called {\it Statistical Parallax} 
and {\it Ordered Motion Parallax}, respectively.

Because IC\,133 is much weaker (due to its far greater distance) than any 
Galactic H$_2$O maser source, its observation requires a much longer time 
baseline.  At the distance of IC\,133 a typical transverse velocity of 30 km 
s$^{-1}$ corresponds to a proper motion of $\approx$ 10 $\mu$as yr$^{-1}$.  
Since the motions of the weaker maser spots are expected to have errors of 
this magnitude, many years are required for a successful observation.  We, 
therefore, undertook a 14 year long VLBI study to track proper motions and 
determine a distance using techniques that have been successfully applied to 
Galactic H$_2$O maser sources.  In $\S$2 we present the details of the VLBI 
observations, calibrations, and results; in $\S$3 we present separate 
millimeter wave observations of the IC\,133 maser region that provide 
independent parameter constraints for our modeling; in $\S$4 we outline the 
proper motion modeling and present results; and in $\S$5 we draw conclusions 
about the importance of our geometrical technique for determining 
extragalactic distances.

\section{22 GHz VLBI Investigations}
\subsection{Observations}
Observations of the H$_2$O maser in IC\,133 were made using the US VLBI 
network and its successor the VLBA, operated by the National Radio Astronomy 
Observatory (NRAO)\footnote
{The National Radio Astronomy Observatory is operated by Associated 
Universities Inc., under cooperative agreement with the National Science 
Foundation.}  
The maser was observed on eight different occasions between 1987 and 2001 
(Table 2).  Approximately 9 hours was spent on-source during Epoch 1, 14 hours 
during each of Epochs 2$-$7, and 1.5 hours during Epoch 8.  In order to 
minimize systematic errors, we kept the arrangement of proper motion 
measurements, the array, the sequence of scans, the source lists, and the LO 
tuning as fixed as possible from epoch to epoch.  
Two epochs, Epoch 8 and the second day of Epoch 1, i.e., June 7, required 
slightly different observational setups, since a second maser, M33.19 
\citep{huc78,ivk74}, was also observed.  In particular, during Epoch 8 and the 
second half of Epoch 1, we switched rapidly between IC\,133, one or more 
calibrators, and M33.19, while in Epochs 2$-$7 we observed 1.75 hour blocks of 
IC\,133, bracketed by calibrators.  Rapidly switched observations were made so 
that the relative angular motions of IC\,133 and M33.19 could be measured and 
an independent distance to M33 obtained.  The results of the rapid switching 
observations of Epoch 8 will be reported by Brunthaler, Reid, \& Falcke (in 
preparation).  The data were correlated at the Mark III Haystack correlator 
(epochs 1$-$5) and at the VLBA correlator (Epochs 6$-$8).  

The Mark III video converters (VCs) or VLBA Baseband converters (BBCs) were 
set so that the full range of emission from IC\,133, as observed in 1987 
April, could be covered \citep{gre90}.  In most cases, we used three slightly 
overlapping 2 MHz ($\approx$ 27 km s$^{-1}$) bands set at center frequencies 
corresponding to the velocities $-$216.28, $-$240.35, and $-$264.43 km 
s$^{-1}$, with respect to the local standard of rest (LSR).  (See Table 3 for 
details.)  In the first two epochs, three of the other VCs were set 
redundantly to protect against recording faults and in Epochs 3$-$7 four of 
the VCs or BBCs were set in a manner so that a multi-band delay analysis could 
be performed (see $\S$2.2.1$-$2 below).  In Epoch 8, a single 8 MHz wide band 
was employed for IC\,133 observations.  We observed left circular polarization 
throughout. 

\subsection{Calibration}
We employed different calibration schemes for epochs 1$-$5 than we did for 
epochs 6$-$8 because we 1) used different arrays for the two groups of epochs 
and 2) took advantage of more sophisticated implementations of various 
processing tools for the latter group of epochs.

\subsubsection{Epochs 1$-$5}
After correlation, the data from Epochs 1$-$5 were translated to the NRAO 
Mark II data format and processed with the spectral-line VLBI software package 
at the Center for Astrophysics (CfA) in Cambridge, MA.  Post-correlation data 
reduction included amplitude, delay, and phase calibrations.  Amplitude 
calibration involved correcting cross power spectra for bandpass response and 
variations in station gain and atmospheric opacity.  Delay calibration refined 
station clocks and drift rates and was based on observations of a few strong 
quasars, including 3C\,345, 3C\,454.3, 0234$+$285, and NRAO150.  Phase 
calibration corrected for variations induced by atmospheric and frequency 
standard fluctuations and was based on a technique called spectral ``phase 
referencing'', whereby the phase of one spectral channel (the ``reference 
feature'') was subtracted from the phases of all other spectral channels.  
See \citet{rhb80} for details.  A small part of the processing was 
accomplished in a non-standard fashion with Haystack Observatory spectral-line 
software and is described below.  

In an effort to reduce delay errors and hence relative position errors of 
maser ``spots'' in synthesis maps, we employed a non-standard method of delay 
calibration for Epochs 3$-$5.  Having set the non-maser VCs in these epochs to 
appropriate values, a multi-band analysis was performed by first running the 
Haystack FRNGE program (in manual phase cal as opposed to extracted phase cal 
mode) on all calibrator scans to determine multi-band delay {\it 
residuals}.  These residuals were then used to correct Haystack {\it a priori} 
single-band clocks, which had different values since they referred to 
different points in the signal path.  After correcting the Haystack {\it a 
priori} clocks, we solved for a single linear drift rate for each station and 
a possible clock jump between the ``two days'' of each epoch.  (Each of the 
three epochs consisted of 17 hours of observations, a 7 hour break, and 
another 17 hours of observations.)  The clock drift rates obtained were then 
applied to CfA single-band data to determine actual clocks.

Clock drift rate errors averaged 0.7, 4.3, and 3.2 ns day$^{-1}$, while clock 
errors averaged 0.4, 2.4, and 1.8 ns, for Epochs 3, 4, and 5, respectively.  
We attribute the higher errors in Epoch 4 to poor weather at NRAO and the 
fact that Haystack did not observe and in Epoch 5 to poor weather at both 
NRAO and Haystack.  In contrast, Epoch 2 drift rate and clock errors were 
$<$ 8.6 ns day$^{-1}$ and $<$4 ns, respectively and Epoch 1 clock errors were 
as high as 22 ns.

\subsubsection{Epochs 6$-$7}
After correlation, the data from Epochs 6$-$7 were imported directly into 
NRAO's Astronomical Image Processing System (AIPS) software package.  
Amplitude, multi-band delay, and phase calibrations were carried out in the 
standard fashion for spectral line VLBI data.  Spectral ``phase referencing'' 
was accomplished by extracting a ``reference feature'' from the main band (the 
$-$216.28 km s$^{-1}$ band), self calibrating to remove residual phase and 
amplitude errors, and applying the solutions to the entire data set.

\subsubsection{Epoch 8}
Epoch 8 data were reduced in the standard fashion until the ``phase 
referencing'' step.  In this epoch {\it absolute phases} were determined, in 
contrast to the {\it relative phases} (relative to a given feature, called the 
``reference feature'') that were determined for the seven previous epochs.  
The quasar, J0137$+$312, was used to accomplish this.  Further details will be 
reported in a future paper (Brunthaler et al., in preparation).

\subsection{Imaging}
In agreement with \citet{gre90,gre93}, we have identified three centers of 
maser emission:  1) an area near the reference feature, which was typically 
located at an LSR velocity of about $-$210 km s$^{-1}$, 2) an area 25 mas east 
and 15 mas south of the reference feature, and 3) an area 290 mas west and 45 
mas south of the reference feature.  In \citet{gre90}, regions 1), 2), and 
3) were called IC\,133 Main, IC\,133 Southeast, and IC\,133 West, 
respectively.  We now refer to 1) and 2) together as IC\,133 Main (since the 
two regions are separated by $\sim$ 0.1 pc and are probably associated with 
the same star-forming region) and 3) as IC\,133 West.

Having detected no new areas of emission in any of the epochs, we made 
three small, high resolution maps in each band.  In particular, we imaged 
(with natural weighting) and CLEANed each spectral channel map in $256\times 
256$ pixel boxes centered on: (0.000,0.000), (0.024,$-$0.015), and 
($-$0.288,$-$0.044) mas.  Because the restoring beams varied by up to a factor 
of two in each direction among the epochs (Table 2), the pixel sizes 
and hence box sizes varied too.  The average box size was $16.4\times 16.4$ 
mas.  Epochs 1$-$5 were imaged and CLEANed with the NRAO AIPS task MX and 
Epochs 6$-$8 were imaged and CLEANed with the AIPS task IMAGR.  The 
(u,v)-data were variance weighted to maximize the sensitivity of maps.

After imaging, selected peaks were fit with a two dimensional Gaussian to 
determine their flux density and {\it x} and {\it y} offsets from the 
reference feature.  Most spurious maser spots were rejected by requiring that 
the following three constraints be met:

1) {Peaks have a flux density of at least 5 times the channel rms (Table 2).}

2) {Peaks have a flux density of at least 1.3 times the absolute value 
of the largest negative peak in the channel map.}

3) {Peaks persist over two or more adjacent channels at approximately 
the same position ($<$ beam).}

The last constraint was relaxed slightly when groupings of related peaks, 
called ``features'', were found to persist over many epochs.  In particular, 
when features persisted over three or more epochs, we accepted single channels 
that met the other two constraints.

Maps from individual epochs were then referred to a common origin by selecting 
a strong and relatively isolated feature that was present during all eight 
epochs (Fig. 1).  The accuracy of the registration varies from epoch to epoch 
and depends on factors such as the reference feature's peak intensity, the 
number of channels that compose the feature, and the presence or absence of 
significant systematic error during the epoch.  See $\S$2.4.1 for details of 
the error model. 

\subsection{Results}

The maser features in IC\,133 Main (Fig. 2) appear to trace out two short 
arcs, one on each edge of an ellipse with major axis $\sim$ 30 mas, minor axis 
$\sim$ 10 mas, and position angle $\sim$ $-$65$^{\circ}$, measured east of 
north.  The redshifted spots tend to be clustered on one side of the ellipse 
and the blueshifted spots on the other side.  Most of the proper motions are 
directed outwards from the center of the ellipse.  In IC\,133 West (Fig. 3), 
features are clustered in two regions: the more densely populated region is 
centered at an offset of $\sim$ ($-$284,$-$39) mas and the less densely 
populated region, consisting of just three spots, is centered at an offset of 
$\sim$ ($-$283.5, $-$45) mas.  The proper motions in IC\,133 West are few and 
appear to be random.

Table 4 shows the peak flux denisty, right ascension and declination offsets, 
and uncertainties for all 81 features in IC\,133 Main and 25 features 
in IC\,133 West.  Proper motions were determined for 16 of the features in 
IC\,133 Main and 6 of the features in IC\,133 West.  Four additional features, 
even though they persisted over three or more epochs (as did the 22 features 
for which proper motions were determined), were deemed too spectrally blended 
for an accurate proper motion determination.  

\subsubsection{Position Errors}
The formal uncertainty in fitted position for unblended maser spots is given 
by $\Delta\theta_{fit} \sim 0.5\theta_b\Delta S/S$, where $\theta_b$, 
$\Delta S$, and $S$ are the synthesized beam size, image rms, and peak 
intensity, respectively \citep{rsm88}.  Since the position angles of the 
restoring beams are all close to zero (Table 2), we substituted the restoring 
beam major and minor axes to obtain uncertainties in right ascension and 
declination, respectively.  As an example, the peak channel of the reference 
feature in Epoch 7 has formal uncertainties of 0.7 $\mu$as and 
2.5 $\mu$as in right ascension and declination offsets, respectively.  Since 
the reference feature is an average of a number of channels, however, its 
formal uncertainty is somewhat smaller.

The uncertainty in the absolute position of the reference feature gives rise 
to one of the largest sources of known systematic error in synthesis maps.  
The position error of a maser spot in synthesis maps, $\Delta\theta_{spot}$, 
is related to the position error of the reference feature, 
$\Delta\theta_{ref}$, by $\Delta\theta_{spot} = (\Delta\nu / \nu) 
\Delta\theta_{ref}$, where $\nu$ is the frequency of the maser spot and 
$\Delta\nu$ its frequency offset from the reference feature \citep{tms01}.  
Since the reference feature is known to an accuracy of $\sim 0\as1$ in each 
coordinate, maser spots offset by 1 MHz (13.5 km s$^{-1}$) from the reference 
feature will be subject to systematic position offset errors of $\sim$ 5 
$\mu$as.  For most epochs, systematic errors arising from the reference 
feature's absolute positional uncertainty are roughly the same for a given 
feature and hence do not strongly affect proper motions.  This is because the 
reference features in those epochs lie within $\sim$ 2 km s$^{-1}$ of one 
another (all close to $-$210 km s$^{-1}$).  In two epochs, however, blending 
near $-$210 km s$^{-1}$ forced us to choose reference features at $-$216.28 km 
s$^{-1}$ (Epoch 3) and $-$220.49 km s$^{-1}$ (Epoch 7).  The total spread in 
reference feature LSR velocities is 12.4 km s$^{-1}$, which could lead to 
errors that differ by as much as 4 $\mu$as from one epoch to another for a 
given feature.

Clock offset errors are another source of systematic error.  In Epoch 1 delay 
errors were found to be 22 ns on VLA baselines and 7 ns on non-VLA baselines.  
Since VLA baselines are among the most sensitive baselines in the experiment, 
systematic errors in synthesis maps will be quite significant for this epoch. 
\citet{gre93} did a Monte Carlo simulation to assess the impact of delay and 
other systematic errors on synthesis maps in Epochs 1 and 2 as a function of 
frequency or LSR velocity.  They found $\sim$0.13 and 0.57 $\mu$as (km 
s$^{-1}$)$^{-1}$ for right ascension and declination errors, respectively 
(Epoch 2) and $\sim$0.59 and 2.5 $\mu$as (km s$^{-1}$)$^{-1}$ for right 
ascension and declination errors, respectively (Epoch 1).  These refer to 
errors within the reference band.  In Epochs 3$-$7, we employed a multi-band 
analysis to correct clock offsets and drift rates and do not believe residual 
errors in these quantities to be significant.

When maser spots are weak the root-sum-square of the individual errors 
described above is a good estimate of the total error.  This is because weak 
spots are signal-to-noise limited and the effects of unknown systematic errors 
do not significantly change the error estimate.  When spots are strong, 
however, the root-sum-square above is almost certainly an underestimate of the 
real error.  Strong spots are dynamic range limited, not signal-to-noise 
limited, which means that unknown systematic errors play a much larger role.  
Because of this unknown and unaccounted for error, we assign a minimum 
position error of 10 $\mu$as in each coordinate.  Note that this refers to 
individual channels and not to features, which are typically composed of 
many channels.

\subsubsection{Proper Motions}
There are two ways to determine proper motions:  

1) {{\it The feature method} where related channels in a given epoch are 
grouped into features and proper motions are found from the best fit to 
feature positions as a function of time.}

2) {{\it The channel method} where individual channels are analyzed for 
motion and a weighted average of all motions among related channels is 
calculated.}  

Since many of our features appear to be spectrally blended, possibly 
due to the amplification of different hyperfine components (see $\S$2.4.3 
below), we prefer the ``channel method'' to determine proper motions.

Table 5 shows the results of the ``channel method'' in both IC\,133 Main and 
IC\,133 West.  We note that our channel-by-channel analysis was complicated by 
the fact that the eight epochs had three different spectral resolutions or 
channel widths: epochs 1$-$5 had a resolution of 0.4808 km s$^{-1}$, epochs 
6$-$7 had a resolution of 0.2106 km s$^{-1}$, and epoch 8 had a resolution of 
0.843 km s$^{-1}$.  In order to get channel motions, we assumed that 
channels were separated by 0.4808 km s$^{-1}$ (because an epoch 1$-$5 data 
point was present for all motions) and if a feature persisted into epochs 
6$-$8, we paired up all smaller (or larger) resolution channels that fell 
within $\pm$0.24 km s$^{-1}$ of the 0.4808 km s$^{-1}$ channel.  
This meant, for example, that a single epoch 1$-$5 channel could be paired up 
with two (and sometimes three) epoch 6$-$7 channels.  We fit all pairings but 
because they were not all independent we adjusted the weighted averages and 
error estimates accordingly.  For example, the right ascension proper motion 
error for Feature 4 in IC\,133 West is given by $[0.5/(5.91)^2 + 0.5/(5.86)^2 
+ 1/(6.40)^2]^{-1/2}$, since the first and second channel motions are not 
independent.

Table 6 shows a comparison between the ``channel'' and ``feature'' methods 
of determining proper motions in IC\,133 Main.  In the ``channel method'', we 
required that two criteria be met for a motion to be retained: 1) a channel 
had to persist over at least three epochs and 2) there had to be at least two 
such channels to average.  In the ``feature method'', we asked that 
individual position errors (i.e., for a feature in a particular epoch) not 
deviate by more than 3-$\sigma$ from the least squares fit to all the epochs.  
This was accomplished by either using the error model as is or by doubling all 
errors in the error model.  Seven of the sixteen motions determined by the 
``feature method'' required a doubling of all position errors and one feature 
could not be fit even then.  Thus, even though the two methods give compatible 
results, the ``channel method'' appears to be better suited to spectral 
blends, since it does not require that we artificially inflate errors in 
order to obtain motions.

\subsubsection{Spectra}
We constructed IC\,133 spectra, total imaged flux density (Jy) as a function 
of LSR velocity, from the fitted peaks in synthesis maps (Fig. 4).  When a 
channel had no peaks that fit the three criteria above, the flux density was 
set to the 5-$\sigma$ noise level for that epoch.  In the first epoch we 
plotted results from the main band (the $-$216.2 km s$^{-1}$ band) only, 
since there were no fitted peaks in the remote bands.  In the other epochs we 
do not display the $-$264.43 band because there were only two believable (and 
weak) fitted maser spots in this band.

Figure 5 shows the time evolution of a spectrally blended feature in IC\,133 
Main.  There are at least two explanations for such blending.  First, it could 
be due to the simple superposition of features.  Second, it could be due 
to amplification of different hyperfine components at different times.  
The $6_{16}-5_{23}$ transition of H$_2$O is split into six hyperfine 
components that span $\sim$ 6 km s$^{-1}$ in LSR velocity \citep{wal84}.  
Under the extreme non-LTE conditions of a typical H$_2$O maser region, 
the relative intensities of the hyperfine components might be expected to vary 
a great deal over time and bear little resemblance to LTE ratios.  
\citet{wal84} did a statistical study of 386 H$_2$O maser features in 
W49N and found that all six hyperfine components were present, not just the 
three strongest ones.  The complicated spectral blend located near 
($-$0.2,$-$1.0) mas (Fig. 5) spans $\sim$ 6 km s$^{-1}$ in LSR 
velocity and varies quite a bit from epoch to epoch.  This variation might be 
caused by different hyperfine components being more strongly amplified in 
different epochs.

\section{Owens Valley MM Array Investigations}
We observed four transitions, CO J=1$\rightarrow$0, CO J=2$\rightarrow$1, HCN 
J=1$\rightarrow$0 and HCO$+$ J=1$\rightarrow$0 with the Owens Valley Radio 
Observatory's Millimeter Array between 1998 April and 1999 February (Table 7). 
The CO J=1$\rightarrow$0 observing band consisted of 128 spectral channels at 
a velocity resolution of 0.65 km s$^{-1}$, the CO J=2$\rightarrow$1 observing 
band consisted of 64 spectral channels at a velocity resolution of 0.65 km 
s$^{-1}$, and the HCN J=1$\rightarrow$0 and HCO$+$ J=1$\rightarrow$0 observing 
bands consisted of 32 spectral channels at velocity resolutions of 0.85 and 
0.84 km s$^{-1}$, respectively.  The CO J=1$\rightarrow$0 band was centered at 
$-$230 km s$^{-1}$ and the others were centered at $-$222 km s$^{-1}$.

Calibration consisted of the following: a flux calibration (using a primary 
calibrator such as Neptune), a passband calibration (using observations of a 
strong quasar such as 2145$+$067), and a gain calibration.  After calibration, 
we combined the (u,v)-data sets from the seven CO J=1$\rightarrow$0 
tracks (low, high, and ultrahigh resolution configurations) and mapped and 
CLEANed all 128 channels with the AIPS task IMAGR using a restoring beam of 
$2\as2\times 1\as8$.  Similarly, we combined the (u,v)-data from the two 
CO J=2$\rightarrow$1 tracks and mapped and CLEANed all 64 channels using a 
restoring beam of $1\as0\times 0\as8$.  After mapping, we found the moments 
in the two images using the AIPS task XMOM.  Figure 6 shows the zeroeth 
moment of the two transitions, superimposed on the same map, and Fig. 7 
shows spectra, constructed by plotting the total imaged flux density within a 
small area as a function of LSR velocity.

\section{Discussion}

There are two techniques to estimate distances from H$_2$O proper motion 
measurements.  The first, {\it Ordered Motion Parallax}, is appropriate for 
sources that can be described by simple models where distance is a parameter, 
e.g., spherically symmetric outflow or symmetrical rotation.  In this 
technique, a dynamical model is fit to observed proper motion and LSR velocity 
distributions.  The distance is one of the parameters of the fit that is 
necessary to couple proper angular motions and linear line of sight 
velocities.  This technique has been used successfully to estimate the 
distance to a number of Galactic star-forming regions including Sgr B2N 
\citep{rsm88}.  A second and simpler technique, {\it Statistical Parallax}, is 
appropriate for sources with isotropic random motions \citep{trw53}.  The 
distance is given by $D$ = $\sigma _R/\sigma _T$, where $\sigma _R$ and 
$\sigma _T$ are the dispersions in radial Doppler motions and in transverse 
angular motions, respectively.  This method has also been used successfully to 
find the distance to a number of Galactic star-forming regions including 
W51\,M \citep{gen81}, W51\,N \citep{sld81}, and Orion-KL \citep{grm81}.  We 
also used it in a preliminary distance estimate to M33 \citep{gre93}. 

\subsection{IC\,133 Main}
\subsubsection{The Model}

The structure and velocities of IC\,133 Main (Fig. 2) suggest that it can be 
modelled as uniform outflow from a central source.  For each maser feature we 
have measured a radial velocity, $v_z$, and the position offsets, $x$ and 
$y$.  In addition, we measured two components of proper motion, $v_x$ and 
$v_y$, for all maser features that persisted over three or more epochs.  Seven 
global parameters are unknown: distance to the source, $D$; the velocity of 
expansion, $V_{exp}$; the velocity in the plane of the sky of the center of 
expansion with respect to the reference feature, $v_{x_0}$ and $v_{y_0}$; the 
radial velocity of the center of expansion, $v_{z_0}$; and the sky position of 
the center of expansion, $x_0$ and $y_0$.  In addition, the positions of the 
masers along the line of sight, $z$, are unknown.  These parameters are 
related by the equations \citep{rsm88}: 
$$v_x = {V_{exp}\over d} {(x-x_0)\over R} + v_{x_0},\eqno(1)$$
$$v_y = {V_{exp}\over d} {(y-y_0)\over R} + v_{y_0},\eqno(2)$$
$$v_z = V_{exp} {z\over R} + v_{z_0},\eqno(3)$$
where $R = [(x-x_0)^2 + (y-y_0)^2 + z^2]^{1/2}$ and $d=D/D_0$, with $D_0$ 
being an {\it a priori} distance.  

Initial estimates of the global parameters and the radial offsets provide a 
starting point to the fitting procedure.  We assume that: 

1) {$D_0$ is 700 kpc.  This number is used solely to convert the units of 
observed proper motion from $\mu$as yr$^{-1}$ to km s$^{-1}$.  Our choice does 
not affect the outcome.}  

2) {$V_{exp}$ and $v_{z_0}$ are constrained as outlined below.}

3) {$v_{x_0}$ and $v_{y_0}$ are the weighted averages of $v_x$ and $v_y$, 
respectively.}

4) {($x_0,y_0$) is (0,0).}

5) {the radial offsets, $z$, are estimated from Eq. (3) and 2) above.}

One additional point deserves special mention.  Our experience with other 
H$_2$O masers such as Sgr\,B2N \citep{rsm88}, as well as residuals to initial 
fits of the IC\,133 velocity field, lead us to expect significant turbulent 
motions intrinsic to the source.  In Sgr\,B2N, these motions are $\sim$ 15 km 
s$^{-1}$.  We adopt this value for IC\,133, which is reasonable given a 
resulting reduced $\chi^2 \sim$ 1 for fits.  Following \citet{rsm88}, we 
assigned weights to the velocity data by combining this turbulent velocity in 
quadrature with the measurement uncertainties.

\subsubsection{Observations Used to Constrain the Solution}

Because of substantial correlations between the expansion velocity 
($V_{exp}$), the systemic velocity ($v_{z_0}$), and the distance (D), we seek 
independent estimates for $V_{exp}$ and $v_{z_0}$ in IC\,133 Main. 

The range of likely expansion velocities in IC\,133 Main can be estimated by 
comparison with the most powerful Galactic H$_2$O masers.  IC\,133 Main 
resembles these sources in a number of ways.  First, it is comparable in 
{\it scaled flux density} to the four most luminous H$_2$O maser sources in 
our Galaxy.  Scaled flux density is the flux density that would be observed if 
all sources were moved to the distance of IC\,133.  Table 8 shows the peak 
flux densities of the four most luminous Galactic H$_2$O masers, scaled
to a distance of 850 kpc, an unweighted average of the 13 M33 distance 
estimates in Table 1.  

Second, the radial velocity span in IC\,133 Main is similar to what one 
would expect to see in the strongest Galactic masers, were they moved to the 
distance of M33.  For example, the range in detectable features for W49\,N 
at 850 kpc is only $\sim$ 14 km s$^{-1}$ \citep{wmg82}.  It is therefore 
likely that IC\,133 has many weak high velocity features spanning decades or 
more in radial velocity that are simply below our detection limit.  

Third, IC\,133 resembles the most powerful Galactic H$_2$O masers both 
spatially and spectrally.  Fig. 6 of \citet{gre90} shows a spot map of W49\,N, 
were it at the distance of M33.  The spatial extent and distribution of W49\,N 
at this distance is very similar to the spatial extent and distribution of 
maser spots in IC\,133 Main.  Spectrally, IC\,133 is skewed, with most of its 
strong features to one side of its peak.  Three of the four Galactic masers in 
Table 8 (including W49\,N) have simliarly skewed spectral distributions. 

Not only does the H$_2$O maser in IC\,133 Main resemble the most powerful 
Galactic H$_2$O masers, but the compact continuum emission associated with 
it does as well.  In particular, \citep{gre90} fit their 15 GHz IC\,133 
continuum data to a model of an optically thin homogeneous spherical H\,II 
region.  For an observed flux density of 5.5 $\pm$ 0.5 mJy, they calculated 
the number of Lyman continuum photons in the H\,II region to be 2.8 $\times$ 
10$^{50}$ s$^{-1}$.  This, they point out, is consistent with the presence of 
several O4 stars.  \citep{gdm78} present VLBI observations of 12 Galactic 
H$_2$O sources in regions of star formation.  They suggest that each ``center 
of H$_2$O activity'' in a source be identified with a young, massive star.  
W49\,N, for example, with the largest number of centers of H$_2$O maser 
activity (10), might then be associated with 10 OB stars.  

The four most powerful Galactic H$_2$O masers have relatively high velocity 
flows, i.e., expansion speeds in the range 40-55 km s$^{-1}$.  It is 
reasonable to assume that IC\,133 does as well, and that its narrow range of 
observed radial velocities is simply a selection effect due to its great 
distance.  Taking an average of the expansion speeds in Table 8 and an error 
equal to the standard deviation, we adopt an expansion velocity for IC\,133 
of 46$\pm$6 km s$^{-1}$.

The range of likely systemic velocities in IC\,133 Main comes from independent 
CO observations.  CO, which is an excellent tracer of compact regions of star 
formation including H$_2$O masers, consistently gives a systemic velocity of 
$-$222 km s$^{-1}$ for the giant molecular cloud associated with the IC\,133 
maser in M33 (Fig. 7).  The systemic velocity error was taken to be the half 
width half maximum of the two spectra discussed in $\S$3.  We note that even 
though the CO J=2$\rightarrow$1 emission is offset by $\sim 1''$ from the 
maser position, it is well within the contours of the CO J=1$\rightarrow$0 
emission (Fig. 6).  We suggest that the CO J=2$\rightarrow$1 image emphasizes 
a local ``hot spot'' in the giant molecular cloud that either 1) the much 
larger beam of the CO J=1$\rightarrow$0 observation ($2\as 2\times 1\as 8$) is 
unable to resolve or 2) is obscured owing to high optical depth in the CO 
J=1$\rightarrow$0 transition.  Finally, internal motions within the cloud are 
expected to be small, meaning that the systemic velocity of the maser should 
not differ significantly from the measured CO velocities.  A BIMA study of 148 
giant molecular clouds in M33 \citep{eng03}, for example, shows that the 
average full width half maximum in velocity within a cloud is only 5 km 
s$^{-1}$.  Thus we adopt the systemic velocity of the IC\,133 Main H$_2$O 
masers to be $-$222 $\pm$ 5 km s$^{-1}$. 

\subsubsection{Fits to the Velocity Field}

Initial fits to the IC\,133 velocity data show that the center of expansion 
is not well determined.  We have found two well defined local minima in the 
$\chi^2$-plane, one at (0.6,-1.8) mas and the other at (26.0,-13.4) mas, 
which lie near the northwest and southeast edges of the maser 
distribution, respectively.  Reduced $\chi^2$ values for the two centers are 
$\approx$ 0.5 and 0.8, respectively.  Fitted distances are comparable at the 
two centers and are $\approx$ 740 kpc.  A third less robust local minimum lies 
near the center of the distribution and gives distances in the range 630$-$875 
kpc.  Reduced $\chi^2$ values for this region are $\approx$ 0.9.  Since the 
northwest local minimum is stable and has the lowest reduced $\chi^2$, we did 
all subsequent fits at this position, but {\it allowed for a 16\% uncertainty 
in distance due to the uncertainty in the location of the center of 
expansion}.  We note that it is not unusual for H$_2$O maser sources to 
have centers of expansion that are not in the middle of the maser 
distributions.  The four best-fitting models for the maser velocity field in 
W49\,N \citep{gmr92} all give centers of expansion significantly offset from 
the center of the distribution.  An even more asymmetric maser lies in 
Sgr\,B2N \citep{rsm88}, whose center of expansion lies {\it outside of} the 
maser distribution.  For the case of IC\,133, we suggest a physical model in 
which the young star that drives the observed outflow is located near the 
front of a dense cloud core.  The redshifted maser emission, which is tightly 
clustered in the sky, represents (constrained) flow into dense ambient 
material.  The ralatively distant blueshifted emission represents flow that 
has ``broken out'' into lower density material.

Due to the substantial correlations between the expansion velocity, the 
systemic velocity, and the distance, we adopted the following analysis 
strategy.  We solved for distances for a grid of fixed expansion and systemic 
velocities.  We then fit a second order polynomial to distance as a function 
of systemic velocity for each value of expansion velocity in the grid (Fig. 
8).  The center grid value, $V_{exp}$ = 46 km s$^{-1}$ and $v_{z_0}$ = $-$222 
km s$^{-1}$, gives us our distance estimate, and its fitting error, the random 
error due to ``noise'' in the proper motions.  One component of systematic 
error arises from the 16\% ``center of expansion error'' mentioned above.  
Further systematic errors come from the variation in distance of grid values 
corresponding to the error bars on $V_{exp}$ and $v_{z_0}$, also discussed 
above.

We find a distance to IC\,133 (and M33) of 
$$ D(kpc) = 800 \pm 12\% \pm 16\% \pm 9\% \pm 4\%.$$
The quoted errors are due to random error, the systematic error arising from 
the uncertainty in the center of expansion, and the systematic errors due to 
lack of knowledge of $V_{exp}$ and $v_{z_0}$, respectively.  Adding the four 
sources of error in quadrature, we obtain 
$$ D(kpc) = 800 \pm 22\% = 800 \pm 180,$$

Note that the method of {\it Statistical Parallax} can not be used to 
estimate the distance to IC\,133 Main, due to the poor apparent distribution 
of masers.  Even a constrained {\it Statistical Parallax}, i.e., a radial 
velocity distribution centered on an observed systemic velocity of $-$222 km 
s$^{-1}$, gives the unrealistically low distance of $\sim$ 200 kpc.

\subsection{IC\,133 West}
The motions in IC\,133 West appear to be random but are too few for a
meaningful application of the method of {\it Statistical Parallax}.

\section{Conclusions}
We have mapped H$_2$O maser emission in IC\,133 over 8 epochs, spanning 14
years.  We detected 106 features and measured motions in two centers of
activity, Main and West.  The structure and dynamics of IC\,133 Main suggest
radial outflow from a central source; the motions in IC\,133 West appear to be
random.  We have modelled the dynamics of IC\,133 Main to estimate a distance
to M33.  In order to constrain parameters that are correlated, we
independently estimated expansion speed from comparison to the strongest
Galactic sources and systemic velocity from CO J=2$\rightarrow$1 and CO
J=1$\rightarrow$0 observations.  We obtained a distance of 800 $\pm$ 180
kpc.  

Although our results with this data set do not give small enough errors 
for comparison with other distances to M33, the technique is potentially 
powerful.  As mentioned above, a variant of {\it Ordered Motion Parallax} 
has been used to obtain a distance to the galaxy NGC\,4258 with a 7\% 
uncertainty.  The ``maser method'' employed in this paper and for NGC\,4258 is 
independent of the calibrations implicit in the distance indicators used in 
Table 1.  The large scatter and small error bars in this table, even for very 
recent estimates, suggest that uncontrolled systematic errors still exist 
despite the application of much effort, which reinforces the need for 
independent, geometric techniques.

\acknowledgments
The authors are grateful for the generous support received for this project
through multiple grants from the Scholarly Studies program of the Smithsonian
Institution.  We would also like to thank Andreas Brunthaler for providing 
one epoch of data (Epoch 8: 2001 March 27).

\clearpage




\clearpage

\begin{figure}
\epsscale{.35}
\plotone{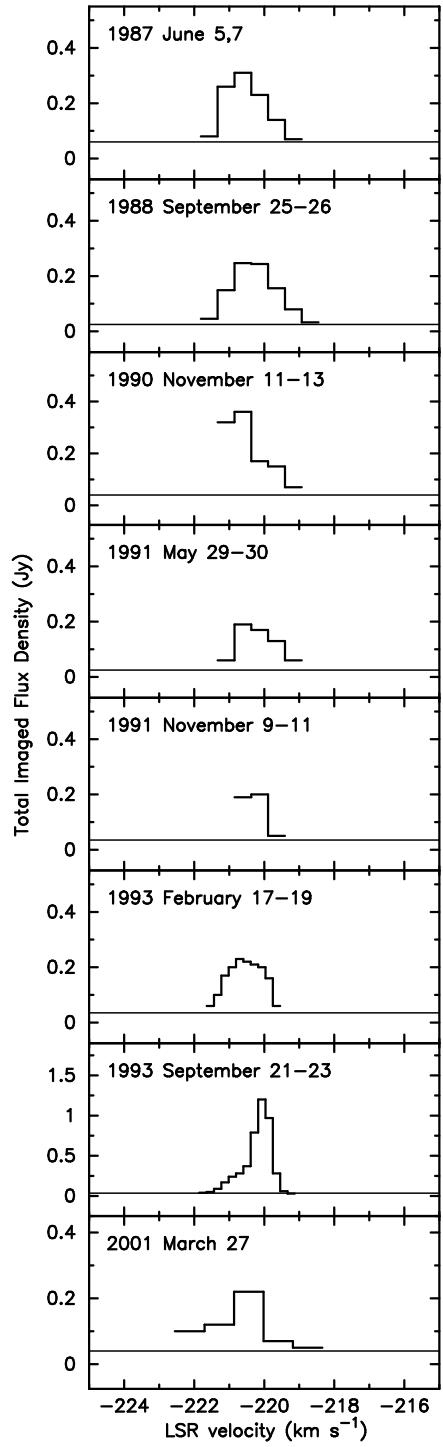}
\caption{Spectra of the maser feature used to register maps for each epoch 
to a common origin.\label{fig1}}
\end{figure}

\clearpage

\begin{figure}
\epsscale{0.80}
\plotone{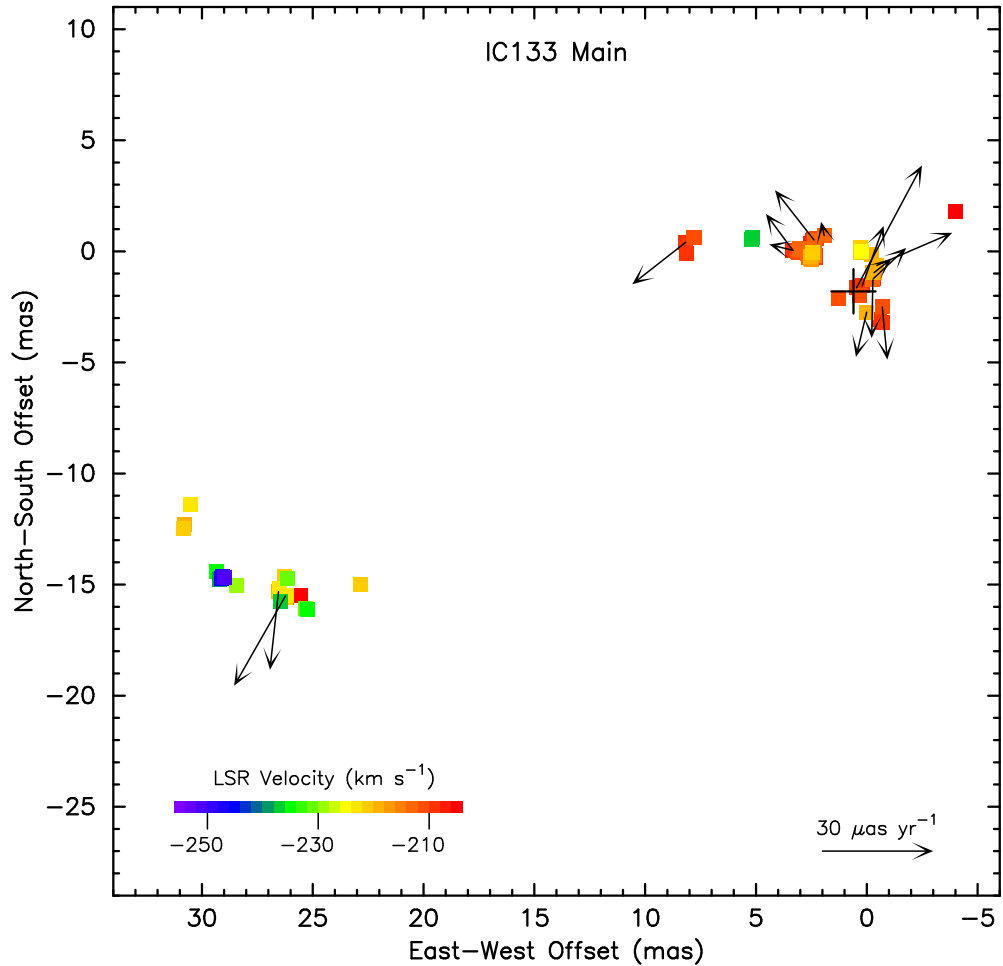}
\caption{Map of features and proper motions in IC\,133 Main.  A constant 
vector has been subtracted from all proper motions such that their weighted 
mean is zero.  Radial velocities are indicated by color.  Note that features 
cluster along two arcs, one on either side of an ellipse with position angle 
$-65^{\circ}$.  Redshifted features tend to lie near the origin, while 
blueshifted features tend to lie to the southeast.  Note that the absence of 
arrows for many features does {\it not} imply that there is no motion in the 
underlying gas.  The cross near the origin marks the location of the center 
of expansion.\label{fig2}}
\end{figure}

\clearpage

\begin{figure}
\epsscale{0.80}
\plotone{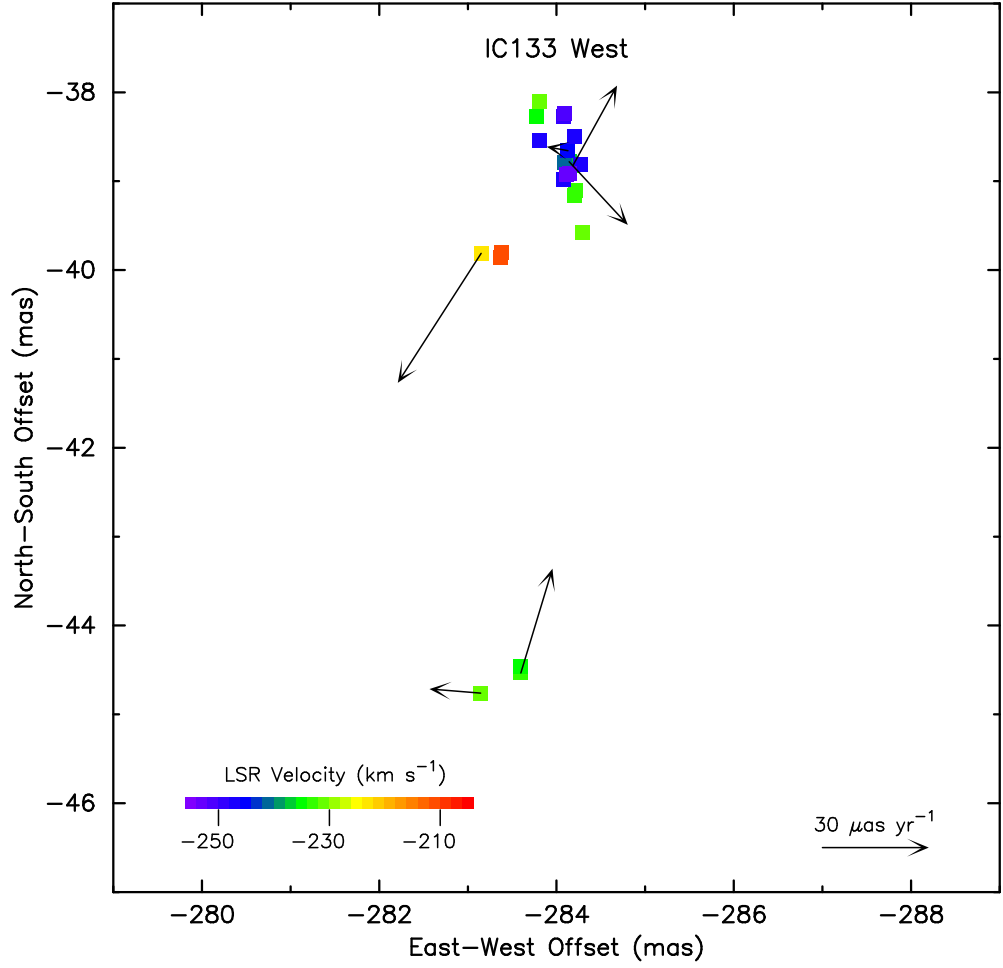}
\caption{Map of features and proper motions in IC\,133 West.  See the 
Fig. 2 caption for details.\label{fig3}}
\end{figure}

\clearpage

\begin{figure}
\epsscale{.70}
\plotone{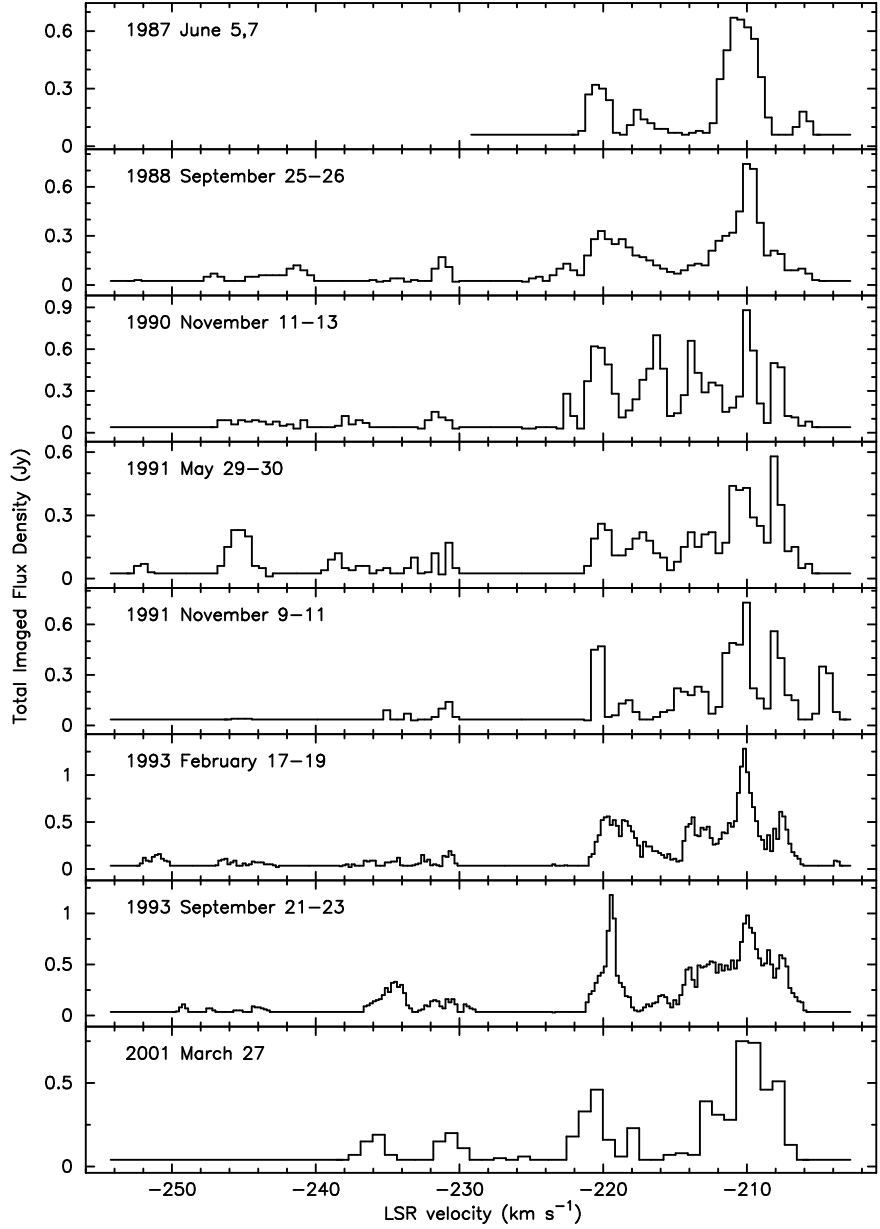}
\caption{Spectra of total imaged power in synthesis maps for each epoch.  
Measured peak flux densities were corrected by the multiplicative factor 
$\theta^2_{meas}/\theta^2_{beam}$, where $\theta^2_{meas}$ is the measured 
peak area and $\theta^2_{beam}$ is the beam area.  No features were found in 
the second band (v$_{LSR} < -230$ km s$^{-1}$) during Epoch 1 (1987 June) and 
only two weak features in the third band (v$_{LSR} < -254$ km s$^{-1}$) for 
any epoch.  We do not display the third band. For channels in which no maser 
emission was detected, the noise level is plotted (5-$\sigma$)\label{fig4}}
\end{figure}

\clearpage

\begin{figure}
\epsscale{.35}
\plotone{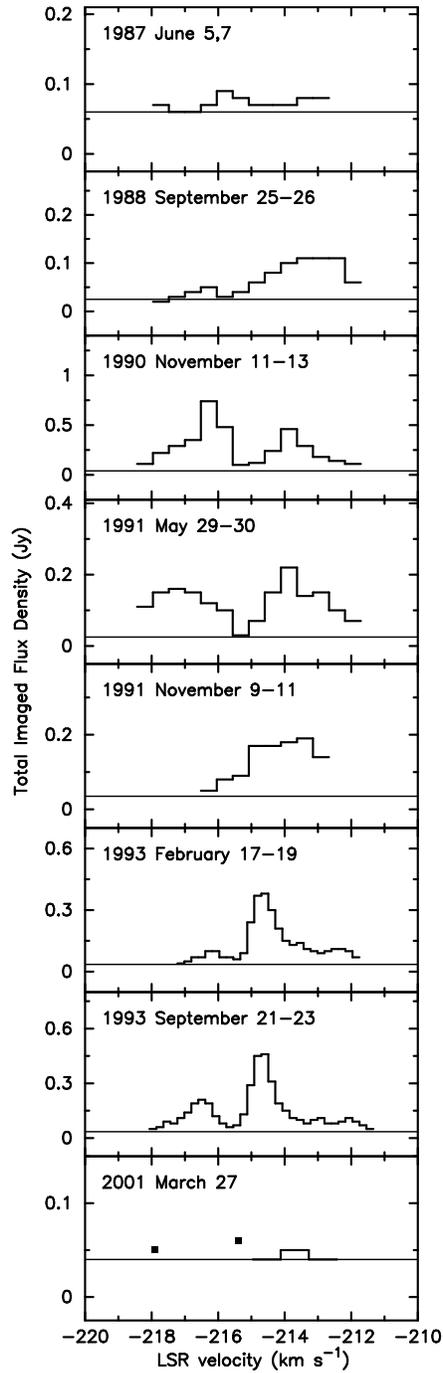}
\caption{Time evolution of a spectrally blended feature in IC\,133 Main.  The 
blending could be due to either the superposition of features or the 
amplification of different hyperfine components. The filled squares in the 
2001 March 27 panel indicate detection of flux in isolated channels.
\label{fig5}}
\end{figure}

\clearpage

\begin{figure}
\epsscale{0.7}
\plotone{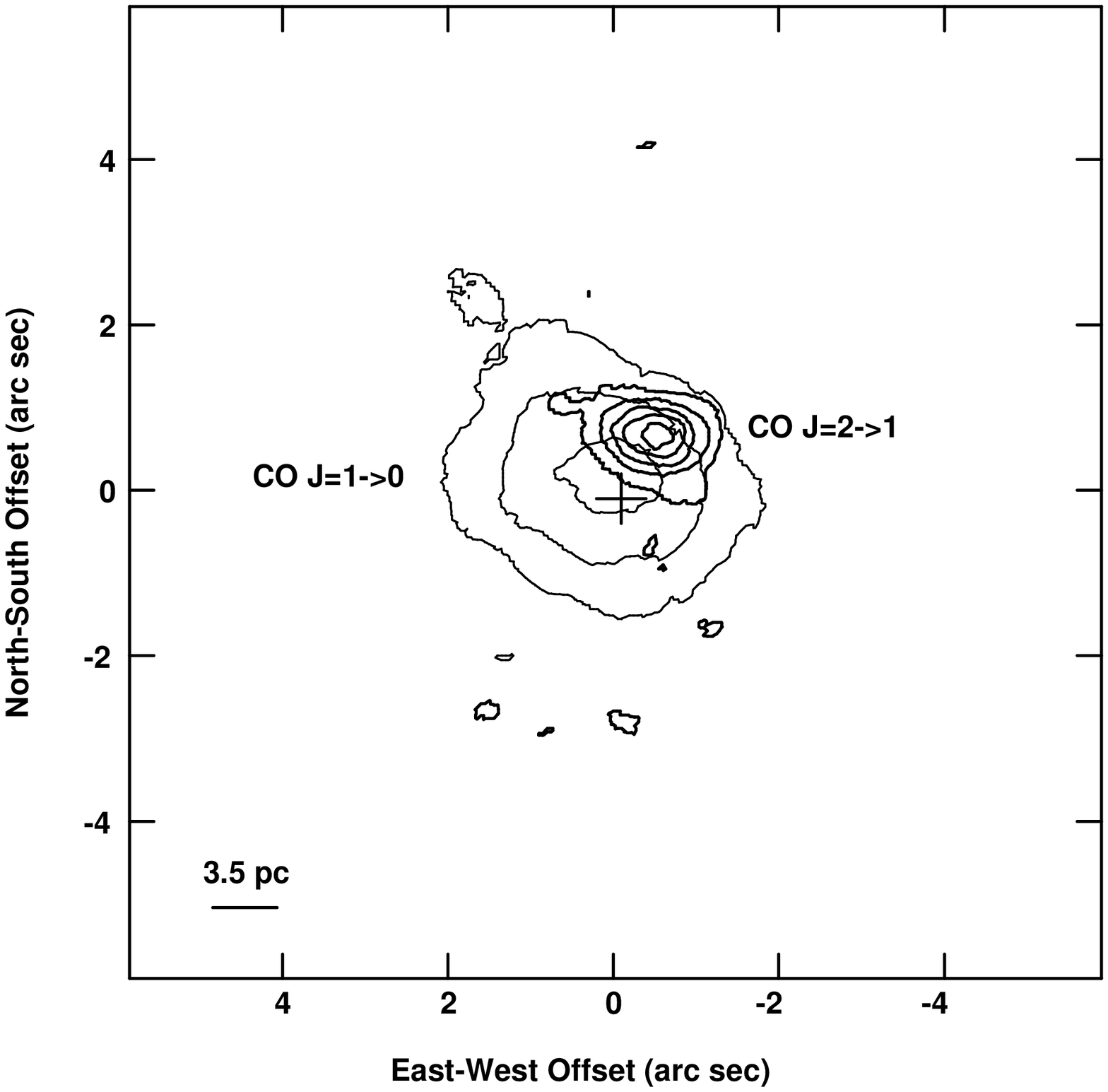}
\caption{Zeroeth moment of the CO J=1$\rightarrow$0 image (light contours) and
the CO J=2$\rightarrow$1 image (heavy contours).  The zeroeth moment is the
flux density summed over all channels.  The restoring beams were $2\as2\times
1\as8$ and $1\as0\times 0\as8$ for the CO J=1$\rightarrow$0 and
CO J=2$\rightarrow$1 transitions, respectively.  Contours are displayed at 10,
30, 50, 70, and 90\% of the CO J=2$\rightarrow$1 maximum, $6.9587\times 10^3$ 
Jy beam$^{-1}\times$ m s$^{-1}$.  We assume a distance of 850 kpc in fixing 
the scale bar.  The cross at the center marks the position of the 
maser.\label{fig6}}
\end{figure}

\clearpage

\begin{figure}
\epsscale{.80}
\plotone{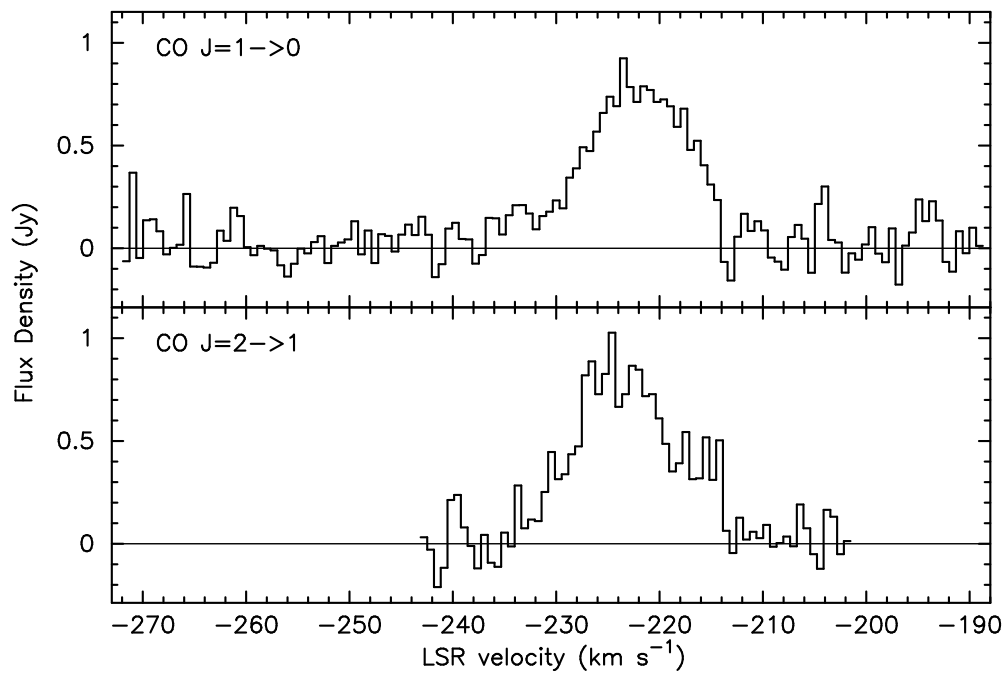}
\caption{Spectra for the CO J=1$\rightarrow$0 (upper) and CO 
J=2$\rightarrow$1 (lower) transitions.  Spectra were constructed by plotting 
the total flux density within a small area as a function of channel.  For 
CO J=1$\rightarrow$0 we summed flux density in a $5\as25\times 5\as45$ area 
centered on ($0\as525,0\as275$) and for CO J=2$\rightarrow$1 in a 
$1\as75\times 1\as5$ area centered on ($-0\as475,0\as55$).  These areas 
correspond to the edges of the 10\% contours of Figure 6.\label{fig7}}
\end{figure}

\clearpage

\begin{figure}
\epsscale{.80}
\plotone{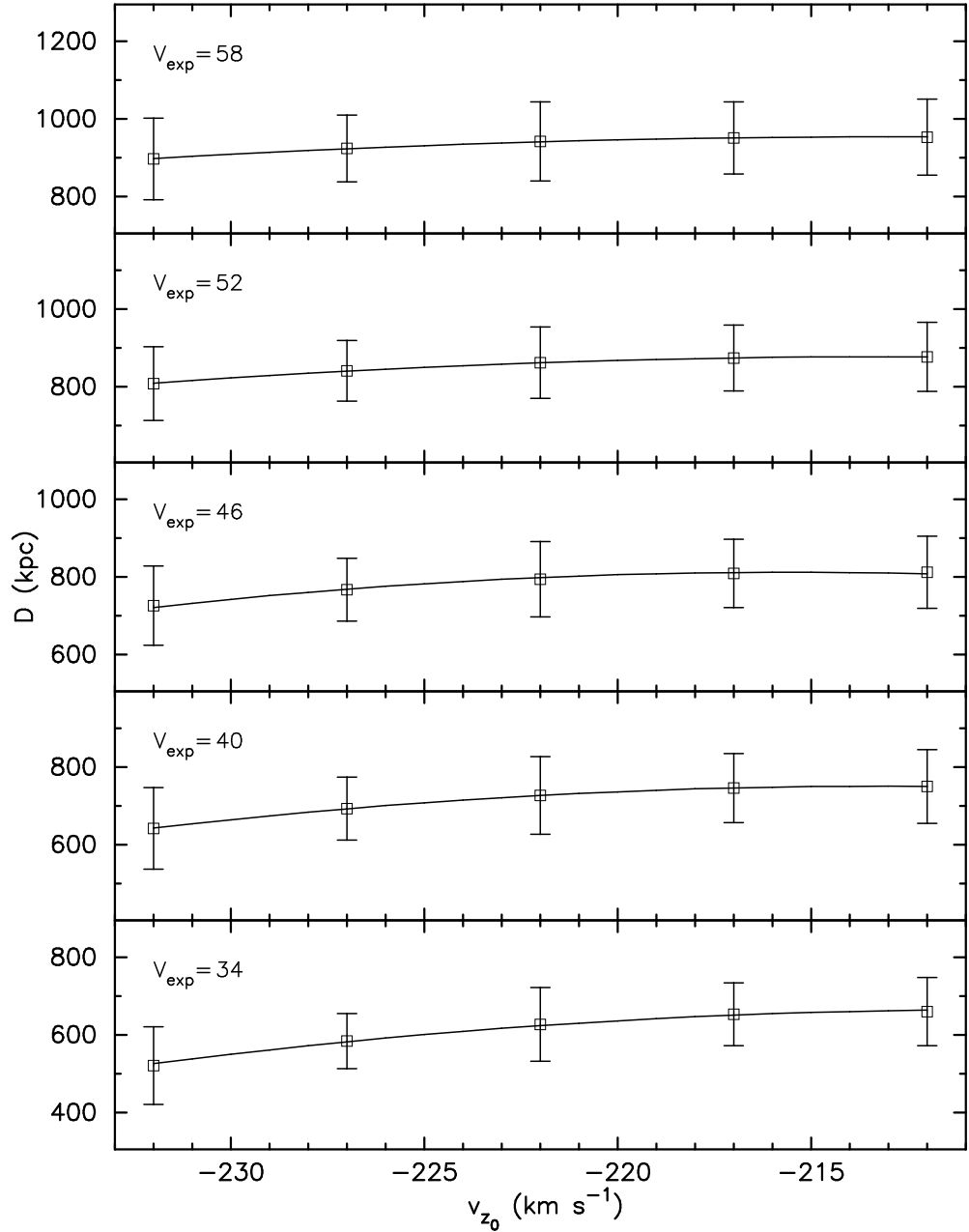}
\caption{Distance ($D$) as a function of systemic velocity ($v_{z_0}$) for 
each of five expansion velocities ($V_{exp}$) in IC\,133 Main.  Error bars 
indicate random fitting errors.  The solid lines are weighted fits, assuming a 
second order polynomial.\label{fig8}}
\end{figure}

\clearpage

\begin{deluxetable}{lll}
\tabletypesize{\scriptsize}
\tablewidth{0pc}
\tablenum{1}
\tablecolumns{3}
\tablecaption{Previous Distance Estimates to M33}

\tablehead{
\colhead{Study} &
\colhead{Method Employed\tablenotemark{a}} &
\colhead{Distance\tablenotemark{b}}\\
&
&
\colhead{(kpc)}
}

\startdata
\citet{kim02}& TRGB             & 916$\pm$17\\
             & RC               & 912$\pm$17\\
\citet{lee02}& Cepheids         & 802$^{+73}_{-67}$\\
\citet{fre01}& Cepheids         & 817$^{+38}_{-37}$\\
\citet{mag00}& Planetary Nebulae& 840$\pm$90\\
\citet{pjc00}& LPVs             & 933$^{+58}_{-54}$\\
\citet{sar00}& HB               & 929$^{+71}_{-66}$\\
\citet{lfm93}& TRGB             & 871$^{+41}_{-39}$\\
\citet{fwm91}& Cepheids         & 847$^{+36}_{-34}$\\
\citet{cs87} & Cepheids         & 692$^{+49}_{-46}$\\
\citet{mou87}& LPVs             & 890$\pm$135\\
\citet{mk86} & TRGB             & 912$^{+135}_{-118}$\\
\citet{dev78}& Cepheids         & 720$\pm$70


\enddata

\tablenotetext{a} {TRGB: tip of the Red Giant branch; RC: the Red Clump; 
LPVs: Long-Period Variables; HB: Horizontal Branch stars}

\tablenotetext{b} {If an actual distance was given, we present it as quoted. 
If only a distance modulus and associated error were given, we converted 
to a linear distance (in kpc) using the formula 
[10$^{(\rm distance\ modulus)/5}$]/100, which resulted in an asymmetric 
distance error.}

\end{deluxetable}


\clearpage

\begin{deluxetable}{lcllcccc}
\tabletypesize{\scriptsize}
\tablewidth{0pc}
\tablenum{2}
\tablecolumns{8}
\tablecaption{VLBI Observations}

\tablehead{
\colhead{Date\tablenotemark{a}} &
\colhead{Epoch} &
\colhead{Antennas\tablenotemark{b}} &
\colhead{Notes} &
\colhead{Major\tablenotemark{c}} &
\colhead{Minor\tablenotemark{c}} &
\colhead{P.A.\tablenotemark{c}} &
\colhead{Noise Level\tablenotemark{d}}\\
&
&
&
&
\colhead{(mas)}&
\colhead{(mas)}&
\colhead{($^{\circ}$)}&
\colhead{(mJy)}
}

\startdata

1987 June 5, 7        & 1& BKGYON        & no data from N;       & 1.25& 0.25& \phantom{,}173  & 60\\
                       &  &               & poor weather at K     &     &     &                 &   \\
1988 September 25$-$26 & 2& BKGYON        &                       & 0.95& 0.26& \phantom{,}166  & 25\\
1990 November 11$-$13  & 3& BKGYON        & no data from N        & 0.75& 0.16& $-$15           & 40\\
1991 May 29$-$30       & 4& BGY VLBA(1) N & poor weather at G     & 0.64& 0.12& \phantom{1}$-$7 & 25\\
1991 November 9$-$11   & 5& BKGY VLBA(1) N& no data from N;       & 0.84& 0.12& $-$13           & 35\\
                       &  &               & poor weather at K,G   &     &     &                 &   \\
1993 February 17$-$19  & 6& BY VLBA(9)    & poor weather at       & 0.65& 0.21& $-$11           & 35\\
                       &  &               & N. Liberty, Hancock   &     &     &                 &   \\
1993 September 21$-$23 & 7& BY VLBA(9)    &                       & 0.85& 0.24& $-$21           & 35\\
2001 March 27          & 8& VLBA          &                       & 0.68& 0.33& \phantom{1}$-$2 & 40


\enddata

\tablenotetext{a} {Epoch 1: $\sim$ 9 hours on-source; Epochs 2$-$7: $\sim$ 14 
hours on-source; Epoch 8: $\sim$ 1.5 hours on-source}

\tablenotetext{b} {B: Max-Planck-Institut f\"ur Radioastronomie 100-m antenna 
in Effelsberg, Germany; K: Haystack Observatory 37-m antenna in Westford, MA; 
G: National Radio Astronomy Observatory (NRAO) 43-m antenna in Green Bank, WV; 
Y: 27$\times$25m phased Very Large Array (VLA) in Socorro, NM; O: Owens Valley 
Radio Observatory 40-m antenna in Big Pine, CA; N: Nobeyama Radio Observatory 
45-m antenna in Nobeyama, Japan; VLBA: Very Long Baseline Array.  VLBA(1) 
refers to the Owens Valley VLBA antenna and VLBA(9) refers to the array minus 
Mauna Kea.}

\tablenotetext{c} {Restoring beam axes and position angle.  Position angles 
are measured east of north.}

\tablenotetext{d} {5-$\sigma$ noise levels in individual channel maps. Channel 
widths vary and are given in Table 3.}

\end{deluxetable}


\clearpage

\begin{deluxetable}{lcccc}
\tabletypesize{\scriptsize}
\tablewidth{0pc}
\tablenum{3}
\tablecolumns{5}
\tablecaption{ Observed Bandpasses \label{}}

\tablehead{
\colhead{Date} &
\colhead{Epoch} &
\colhead{Band center V$_{LSR}$} &
\colhead{Channel separation} &
\colhead{Bandwidth}\\
&
&
\colhead{(km s$^{-1}$)} &
\colhead{(km s$^{-1}$)} &
\colhead{(km s$^{-1}$)}
}

\startdata

1987 June 5$-$7        & 1& $-$216.2\phantom{8}, $-$240.2\phantom{5}, $-$277.8\phantom{3}& 0.4808           & $-$201 to $-$256, $-$262 to $-$293\\
1988 September 25$-$26 & 2&           $-$216.28,           $-$240.35,           $-$264.43& 0.4808           & $-$201 to $-$280\\
1990 November 11$-$13  & 3&           $-$216.28,           $-$240.35,           $-$264.43& 0.4808           & $-$201 to $-$280\\
1991 May 29$-$30       & 4&           $-$216.28,           $-$240.35,           $-$264.43& 0.4808           & $-$201 to $-$280\\
1991 November 9$-$11   & 5&           $-$216.28,           $-$240.35,           $-$264.43& 0.4808           & $-$201 to $-$280\\
1993 February 17$-$19  & 6&           $-$216.28,           $-$240.35,           $-$264.43& 0.2106           & $-$203 to $-$278\\
1993 September 21$-$23 & 7&           $-$216.28,           $-$240.35,           $-$264.43& 0.2106           & $-$203 to $-$278\\
2001 March 27          & 8&                                $-$233.93                     & 0.843\phantom{6} & $-$180 to $-$288


\enddata

\end{deluxetable}


\clearpage

\begin{deluxetable}{ccrrrrc}
\tabletypesize{\scriptsize}
\tablewidth{0pc}
\tablenum{4}
\tablecolumns{7}
\tablecaption{Maser Feature Positions}

\tablehead{
\colhead{V$_{LSR}$\tablenotemark{a}}& \colhead{Peak F$_{\nu}$}& 
\colhead{$\Delta x$\tablenotemark{b}}& \colhead{$\sigma_x$\tablenotemark{b}}& 
\colhead{$\Delta y$\tablenotemark{b}}& \colhead{$\sigma_y$\tablenotemark{b}}& 
\colhead{Epochs}\\
\colhead{(km s$^{-1}$)}& \colhead{(Jy)}& 
\colhead{(mas)}& \colhead{(mas)}& 
\colhead{(mas)}& \colhead{(mas)}& 
\colhead{Observed}
}

\startdata
IC\,133 Main:& &           &         &           &         &   \\
         &     &           &         &           &         &   \\
$-$204.38& 0.04&   $-$3.984&    0.017&      1.799&    0.053&  6\\
$-$204.49& 0.29&     25.538&    0.010&  $-$15.480&    0.049&  5\\
$-$206.17& 0.17&      3.332&    0.005&      0.048&    0.010&  1234\tablenotemark{c}\\
$-$207.62& 0.58&      3.329&    0.002&      0.043&    0.003&  234567\tablenotemark{c}\\
$-$207.80& 0.16&      2.331&    0.013&   $-$0.171&    0.027&  8\\
$-$207.83& 0.09&      2.522&    0.010&      0.374&    0.043&  3\\
$-$207.86& 0.05&      2.437&    0.011&      0.488&    0.041&  2\\
$-$208.18& 0.12&      0.211&    0.008&   $-$1.559&    0.019&  678\tablenotemark{c}\\
$-$208.20& 0.06&   $-$0.730&    0.011&   $-$3.211&    0.042&  2\\
$-$208.22& 0.05&      3.309&    0.026&      0.033&    0.054&  8\\
$-$208.70& 0.09&      8.164&    0.005&      0.412&    0.014&  45678\tablenotemark{c}\\
$-$208.82& 0.06&   $-$0.300&    0.006&   $-$1.223&    0.019&  267\tablenotemark{c}\\
$-$208.82& 0.25&      2.384&    0.004&      0.518&    0.013&  12367\tablenotemark{c}\\
$-$208.86& 0.06&   $-$0.574&    0.020&   $-$3.057&    0.070&  7\\
$-$208.91& 0.05&   $-$0.590&    0.017&   $-$3.065&    0.053&  6\\
$-$209.30& 0.13&      0.462&    0.004&   $-$1.643&    0.010&  13456\tablenotemark{c}\\
$-$209.49& 0.07&      2.299&    0.028&   $-$0.288&    0.139&  1\\
$-$209.52& 0.07&      8.131&    0.026&   $-$0.080&    0.127&  1\\
$-$210.01& 0.08&      7.832&    0.010&      0.616&    0.027&  6\\
$-$210.01& 0.10&      7.769&    0.010&      0.622&    0.032&  7\\
$-$210.11& 0.10&      2.257&    0.010&      0.597&    0.027&  6\\
$-$210.20& 0.27&      1.259&    0.019&   $-$2.100&    0.039&  8\\
$-$210.22& 0.16&      3.091&    0.010&   $-$0.057&    0.027&  5\\
$-$210.51& 0.15&      3.030&    0.010&      0.008&    0.027&  6\\
$-$210.75& 1.16&      2.560&    0.002&   $-$0.257&    0.002&  12345678\tablenotemark{c}\\
$-$210.99& 0.40&      0.311&    0.006&   $-$1.982&    0.009&  345\\
$-$211.23& 0.04&      3.052&    0.011&   $-$0.059&    0.074&  5\\
$-$211.70& 0.06&      3.140&    0.022&   $-$0.016&    0.108&  1\\
$-$211.95& 0.12&   $-$0.699&    0.003&   $-$2.508&    0.010&  123678\tablenotemark{c}\\
$-$212.24& 0.10&      2.978&    0.016&      0.065&    0.056&  7\\
$-$212.32& 0.07&      2.357&    0.033&      0.487&    0.068&  8\\
$-$212.34& 0.15&      2.638&    0.013&   $-$0.262&    0.060&  3\\
$-$212.61& 0.10&      2.599&    0.010&   $-$0.234&    0.027&  4\\
$-$212.66& 0.24&      2.478&    0.010&   $-$0.236&    0.013&  7\\
$-$212.84& 0.14&      2.506&    0.011&   $-$0.272&    0.035&  6\\
$-$213.21& 0.05&      3.045&    0.017&      0.126&    0.053&  6\\
$-$213.39& 0.12&      1.923&    0.005&      0.710&    0.014&  3567\tablenotemark{c}\\
$-$213.40& 0.46&   $-$0.276&    0.002&   $-$1.285&    0.003&  12345678\tablenotemark{c}\\
$-$213.49& 0.22&      2.558&    0.013&   $-$0.178&    0.047&  7\\
$-$213.64& 0.08&      2.672&    0.014&   $-$0.107&    0.042&  6\\
$-$213.69& 0.09&      2.364&    0.020&   $-$0.087&    0.070&  7\\
$-$213.71& 0.07&     26.204&    0.023&  $-$15.572&    0.071&  6\\
$-$215.42& 0.05&      2.349&    0.020&   $-$0.198&    0.070&  7\\
$-$216.46& 0.07&      2.499&    0.020&   $-$0.364&    0.073&  2\\
$-$216.52& 0.74&   $-$0.330&    0.002&   $-$0.957&    0.003&  12345678\tablenotemark{c}\\
$-$217.24& 0.05&      2.528&    0.017&   $-$0.310&    0.053&  6\\
$-$217.73& 0.33&     26.235&    0.003&  $-$15.523&    0.007&  123456\tablenotemark{c}\\
$-$218.05& 0.04&   $-$0.250&    0.023&   $-$0.966&    0.071&  6\\
$-$218.44& 0.04&     22.846&    0.011&  $-$14.983&    0.074&  5\\
$-$218.45& 0.11&      0.002&    0.004&   $-$2.737&    0.015&  234567\tablenotemark{c}\\
$-$219.10& 0.10&   $-$0.203&    0.013&   $-$0.130&    0.047&  7\\
$-$219.65& 0.09&   $-$0.450&    0.007&   $-$0.631&    0.026&  1367\tablenotemark{c}\\
$-$219.78& 0.13&      2.522&    0.012&   $-$0.236&    0.036&  6\\
$-$219.81& 0.04&     30.804&    0.015&  $-$12.306&    0.080&  4\\
$-$219.89& 0.22&     26.548&    0.005&  $-$15.322&    0.012&  2346\tablenotemark{c}\\
$-$220.11& 0.14&      0.298&    0.017&   $-$0.064&    0.053&  6\\
$-$220.44& 0.05&     22.832&    0.026&  $-$14.985&    0.054&  8\\
$-$220.57& 0.03&      0.302&    0.025&      0.184&    0.092&  2\\
$-$220.58& 0.04&     26.229&    0.023&  $-$15.595&    0.071&  6\\
$-$220.61& 0.20&     26.271&    0.013&  $-$14.662&    0.026&  567\\
$-$220.86& 0.04&     30.850&    0.033&  $-$12.472&    0.068&  8\\
$-$221.34& 0.06&      2.431&    0.026&   $-$0.041&    0.054&  8\\
$-$222.45& 0.06&     30.499&    0.022&  $-$11.371&    0.080&  2\\
$-$222.79& 0.02&     26.192&    0.023&  $-$15.505&    0.083&  2\\
$-$222.81& 0.03&     26.487&    0.022&  $-$15.177&    0.081&  2\\
$-$224.86& 0.04&      0.222&    0.025&   $-$0.014&    0.088&  2\\
$-$229.24& 0.04&     28.440&    0.022&  $-$15.037&    0.079&  2\\
$-$230.20& 0.08&     25.326&    0.016&  $-$16.092&    0.034&  8\\
$-$231.03& 0.09&     26.150&    0.015&  $-$14.739&    0.030&  8\\
$-$234.61& 0.11&      5.170&    0.010&      0.613&    0.032&  7\\
$-$235.48& 0.06&     29.358&    0.027&  $-$14.427&    0.093&  7\\
$-$235.66& 0.09&     25.248&    0.026&  $-$16.107&    0.054&  8\\
$-$236.10& 0.06&     26.453&    0.026&  $-$15.771&    0.054&  8\\
$-$236.13& 0.13&      5.177&    0.017&      0.607&    0.056&  7\\
$-$236.98& 0.05&      5.196&    0.019&      0.541&    0.054&  6\\
$-$238.61& 0.05&     29.217&    0.014&  $-$14.775&    0.054&  4\\
$-$243.48& 0.03&     29.085&    0.019&  $-$14.649&    0.061&  2\\
$-$245.81& 0.09&     29.131&    0.016&  $-$14.743&    0.055&  4\\
$-$246.01& 0.05&     29.056&    0.021&  $-$14.642&    0.071&  7\\
$-$246.35& 0.04&     29.045&    0.020&  $-$14.685&    0.054&  6\\
$-$251.45& 0.08&     29.001&    0.021&  $-$14.678&    0.054&  6\\
         &     &           &         &           &         &   \\
IC\,133 West:& &           &         &           &         &   \\
         &     &           &         &           &         &   \\
$-$210.92& 0.09& $-$283.375&    0.015&  $-$39.800&    0.037&  6\\
$-$211.13& 0.19& $-$283.374&    0.016&  $-$39.856&    0.048&  7\\
$-$222.30& 0.14& $-$283.149&    0.010&  $-$39.812&    0.027&  123\tablenotemark{c}\\
$-$230.97& 0.17& $-$283.145&    0.003&  $-$44.761&    0.008&  2367\tablenotemark{c}\\
$-$230.97& 0.16& $-$284.288&    0.006&  $-$39.573&    0.011&  245\\
$-$231.46& 0.03& $-$283.810&    0.018&  $-$38.104&    0.066&  2\\
$-$231.69& 0.11& $-$284.129&    0.005&  $-$38.694&    0.013&  3456\\
$-$232.21& 0.06& $-$284.207&    0.020&  $-$39.158&    0.070&  7\\
$-$232.33& 0.05& $-$284.209&    0.023&  $-$39.099&    0.071&  6\\
$-$233.37& 0.10& $-$283.597&    0.005&  $-$44.535&    0.013&  24567\tablenotemark{c}\\
$-$234.35& 0.03& $-$283.771&    0.019&  $-$38.272&    0.070&  2\\
$-$235.30& 0.10& $-$284.184&    0.007&  $-$38.826&    0.025&  467\tablenotemark{c}\\
$-$235.81& 0.03& $-$283.596&    0.020&  $-$44.461&    0.073&  2\\
$-$237.47& 0.11& $-$284.142&    0.003&  $-$38.781&    0.011&  346\tablenotemark{c}\\
$-$241.37& 0.07& $-$284.085&    0.010&  $-$38.786&    0.043&  3\\
$-$241.51& 0.09& $-$284.114&    0.011&  $-$38.873&    0.041&  2\\
$-$244.44& 0.20& $-$284.129&    0.002&  $-$38.657&    0.005&  234567\tablenotemark{c}\\
$-$246.86& 0.06& $-$284.203&    0.017&  $-$38.491&    0.053&  6\\
$-$246.87& 0.02& $-$283.811&    0.022&  $-$38.538&    0.080&  2\\
$-$247.15& 0.03& $-$284.083&    0.018&  $-$38.983&    0.065&  2\\
$-$247.94& 0.05& $-$284.274&    0.016&  $-$38.814&    0.056&  7\\
$-$249.89& 0.09& $-$284.080&    0.020&  $-$38.271&    0.070&  7\\
$-$251.90& 0.09& $-$284.087&    0.014&  $-$38.235&    0.043&  6\\
$-$252.01& 0.07& $-$284.152&    0.010&  $-$38.911&    0.027&  4\\
$-$253.45& 0.07& $-$284.112&    0.029&  $-$38.926&    0.103&  2


\enddata

\tablenotetext{a} {Average of channel velocities.}

\tablenotetext{b} {East-west and north-south offsets ($\Delta x$ and $\Delta 
y$, respectively) and uncertainties ($\sigma_x$ and $\sigma_y$, respectively), 
measured with respect to the reference maser feature at the epoch of 
observation or at epoch 4 if the feature was observed for more than one epoch. 
If a multi epoch feature was not observed in epoch 4, it was still referred 
to epoch 4.}

\tablenotetext{c} {Proper motions are listed in Table 4.}

\end{deluxetable}


\clearpage

\begin{deluxetable}{llrrrrrrrr}
\tabletypesize{\scriptsize}
\tablewidth{0pc}
\tablenum{5}
\tablecolumns{10}
\tablecaption{Maser spot positions and proper motions}

\tablehead{
\colhead{V$_{LSR}$\tablenotemark{a}}&
\colhead{Epochs\tablenotemark{b}}&
\multicolumn{4}{c}{Position Offsets}& \multicolumn{4}{c}{Proper Motions}\\
\colhead{(km s$^{-1}$)}&& 
\multicolumn{4}{c}{$\hrulefill$}& \multicolumn{4}{c}{$\hrulefill$}\\
&&
\colhead{$\Delta x$\tablenotemark{c}}& \colhead{$\sigma_x$\tablenotemark{c}}&
\colhead{$\Delta y$\tablenotemark{c}}& \colhead{$\sigma_y$\tablenotemark{c}}&
\colhead{$\Delta v_x$\tablenotemark{d}}& \colhead{$\sigma _{v_x}$\tablenotemark{d}}&
\colhead{$\Delta v_y$\tablenotemark{d}}& \colhead{$\sigma _{v_y}$\tablenotemark{d}}\\
& & 
\multicolumn{4}{c}{(mas)}& \multicolumn{4}{c}{($\mu$as yr$^{-1}$)}
}

\startdata
\phantom{ IC 133}IC\,133 Main:   &             &           &      &            &      &          &       &          &      \\
                                 &             &           &      &            &      &          &       &          &      \\
 Feature 1                       &             &           &      &            &      &          &       &          &      \\
$-$205.69                        & 1234        &      3.339& 0.008&       0.030& 0.021&   $-$0.53&   3.40&     10.67& 11.37\\
$-$206.17                        & 1234        &      3.329& 0.008&       0.051& 0.028&   $-$2.61&   3.17&     14.55& 11.51\\
$-$206.66                        & 124         &      3.326& 0.010&       0.054& 0.012&      1.68&   3.91&      7.99& 10.22\\
 Weighted average                &             &      3.332& 0.005&       0.048& 0.010&   $-$0.78&   1.99&     10.82&  6.34\\
                                 &             &           &      &            &      &          &       &          &      \\
 Feature 2                       &             &           &      &            &      &          &       &          &      \\
$-$206.66, $-$206.60             & 35, 7       &      3.344& 0.007&       0.007& 0.021&  $-$32.04&   8.96&     56.28& 29.22\\
$-$206.66, $-$206.81             & 35, 67      &      3.345& 0.007&       0.002& 0.019&  $-$25.32&   6.08&     40.51& 18.19\\
$-$207.14, $-$207.02             & 2345, 67    &      3.341& 0.005&       0.045& 0.011&     10.83&   3.25&  $-$14.06&  9.94\\
$-$207.14, $-$207.23             & 2345, 67    &      3.331& 0.004&       0.048& 0.010&      1.36&   2.71&   $-$7.67&  8.36\\
$-$207.62, $-$207.44             & 2345, 67    &      3.323& 0.004&       0.050& 0.006&   $-$1.67&   2.52&      6.39&  4.80\\
$-$207.62, $-$207.65             & 2345, 67    &      3.326& 0.004&       0.047& 0.005&      0.52&   2.52&      2.08&  4.02\\
$-$207.62, $-$207.86             & 2345, 67    &      3.326& 0.004&       0.047& 0.005&      0.16&   2.52&      1.51&  3.70\\
$-$208.10, $-$207.86             & 2345, 67    &      3.327& 0.004&       0.040& 0.005&   $-$1.59&   2.52&      4.50&  3.84\\
$-$208.10, $-$208.07             & 2345, 67    &      3.325& 0.004&       0.040& 0.005&   $-$3.59&   2.52&      4.22&  3.73\\
$-$208.10, $-$208.28             & 2345, 67    &      3.322& 0.004&       0.039& 0.005&   $-$5.83&   2.52&      3.85&  3.66\\
$-$208.58, $-$208.49             & 45, 67      &      3.327& 0.008&       0.053& 0.024&   $-$9.48&   5.34&   $-$3.75& 13.73\\
$-$208.58, $-$208.70             & 45, 67      &      3.329& 0.008&       0.046& 0.026&   $-$5.98&   6.60&  $-$35.35& 20.80\\
 Weighted average                &             &      3.329& 0.002&       0.043& 0.003&   $-$1.67&   1.44&      2.60&  2.59\\
                                 &             &           &      &            &      &          &       &          &      \\
 Feature 3                       &             &           &      &            &      &          &       &          &      \\
$-$207.86, $-$207.80             & 67, 8       &      0.210& 0.013&    $-$1.519& 0.041&  $-$16.02&   2.00&     14.67&  5.22\\
$-$208.49, $-$208.64             & 67, 8       &      0.211& 0.010&    $-$1.569& 0.021&  $-$11.46&   2.17&     19.08&  4.45\\
 Weighted average                &             &      0.211& 0.008&    $-$1.559& 0.019&  $-$13.93&   1.47&     17.22&  3.39\\
                                 &             &           &      &            &      &          &       &          &      \\
\tablebreak
 Feature 4                       &             &           &      &            &      &          &       &          &      \\
$-$207.86, $-$207.80             & 67, 8       &      8.211& 0.013&       0.458& 0.036&      4.24&   4.29&  $-$13.61&  9.20\\
$-$209.06, $-$208.91             & 45, 7       &      8.166& 0.008&       0.382& 0.018&     12.86&   9.59&     41.01& 30.63\\
$-$209.06, $-$209.12             & 45, 67      &      8.165& 0.008&       0.384& 0.018&     12.97&   6.60&   $-$2.68& 18.36\\
$-$209.54, $-$209.48             & 467, 8      &      8.144& 0.008&       0.465& 0.031&      6.46&   2.23&  $-$10.37&  5.61\\
 Weighted average                &             &      8.164& 0.005&       0.412& 0.014&      6.42&   1.92&  $-$10.36&  4.68\\
                                 &             &           &      &            &      &          &       &          &      \\
 Feature 5                       &             &           &      &            &      &          &       &          &      \\
$-$208.58, $-$208.70             & 2, 67       &   $-$0.307& 0.009&    $-$1.216& 0.030&   $-$12.76&  3.97&      4.32& 13.91\\
$-$209.06, $-$208.91             & 2, 67       &   $-$0.291& 0.007&    $-$1.242& 0.025&   $-$18.72&  3.26&      8.39& 11.50\\
$-$209.06, $-$209.12             & 2, 67       &   $-$0.299& 0.008&    $-$1.213& 0.026&   $-$21.52&  3.50&     17.86& 12.17\\
 Weighted average                &             &   $-$0.300& 0.006&    $-$1.223& 0.019&   $-$16.98&  2.57&      9.28&  9.01\\
                                 &             &           &      &            &      &           &      &          &      \\
 Feature 6                       &             &           &      &            &      &           &      &          &      \\
$-$208.58, $-$208.70             & 123, 6      &      2.401& 0.009&       0.512& 0.030&       3.92&  3.25&     12.72& 12.28\\
$-$209.06, $-$208.91             & 123, 67     &      2.387& 0.005&       0.507& 0.016&       5.21&  1.93&     10.28&  6.11\\
$-$209.06, $-$209.12             & 123, 67     &      2.371& 0.005&       0.530& 0.014&    $-$0.99&  1.84&     18.59&  5.61\\
 Weighted average                &             &      2.384& 0.004&       0.518& 0.013&       2.45&  1.63&     14.41&  5.28\\
                                 &             &           &      &            &      &           &      &          &      \\
 Feature 7                       &             &           &      &            &      &           &      &          &      \\
$-$209.06                        & 345         &      0.474& 0.006&    $-$1.641& 0.012&   $-$63.65& 14.13&     99.73& 37.88\\
$-$209.54                        & 1346        &      0.451& 0.006&    $-$1.649& 0.019&   $-$22.79&  3.98&     21.42& 16.61\\
 Weighted average                &             &      0.462& 0.004&    $-$1.643& 0.010&   $-$25.79&  3.83&     34.05& 15.21\\
                                 &             &           &      &            &      &           &      &          &      \\
 Feature 8                       &             &           &      &            &      &           &      &          &      \\
$-$209.06, $-$209.12             & 2, 67       &      2.532& 0.007&    $-$0.233& 0.010&   $-$18.59&  2.72&      5.45&  3.97\\
$-$209.54, $-$209.33, $-$209.48  & 1234, 67, 8 &      2.552& 0.004&    $-$0.247& 0.006&   $-$23.57&  1.12&      4.17&  2.08\\
$-$209.54, $-$209.48             & 123467, 8   &      2.554& 0.004&    $-$0.257& 0.005&   $-$23.21&  1.12&      1.61&  1.98\\
$-$209.54, $-$209.75, $-$209.48  & 1234, 67, 8 &      2.552& 0.004&    $-$0.252& 0.005&   $-$23.57&  1.12&      3.08&  1.94\\
$-$210.02, $-$209.96             & 12345, 67   &      2.558& 0.004&    $-$0.274& 0.004&   $-$25.12&  1.80&      1.88&  2.37\\
$-$210.02, $-$210.18             & 12345, 67   &      2.558& 0.004&    $-$0.276& 0.004&   $-$25.12&  1.80&      0.65&  2.37\\
$-$210.51, $-$210.39, $-$210.33  & 12345, 67, 8&      2.562& 0.004&    $-$0.254& 0.005&   $-$33.00&  0.91&      6.05&  1.21\\
$-$210.51, $-$210.60, $-$210.33  & 12345, 67, 8&      2.564& 0.004&    $-$0.252& 0.005&   $-$32.79&  0.91&      6.23&  1.21\\
$-$210.99, $-$210.81, $-$211.17  & 123, 67, 8  &      2.575& 0.004&    $-$0.247& 0.006&   $-$32.91&  0.91&      4.77&  1.52\\
$-$210.99, $-$211.02, $-$211.17  & 123, 67, 8  &      2.576& 0.004&    $-$0.249& 0.006&   $-$32.88&  0.91&      4.55&  1.52\\
$-$210.99, $-$211.23, $-$211.17  & 123, 67, 8  &      2.576& 0.004&    $-$0.247& 0.006&   $-$32.83&  0.91&      4.73&  1.52\\
$-$211.47, $-$211.23             & 12, 67      &      2.576& 0.005&    $-$0.247& 0.006&   $-$37.15&  1.84&   $-$3.66&  2.61\\
$-$211.47, $-$211.44             & 12, 67      &      2.578& 0.005&    $-$0.247& 0.006&   $-$36.44&  1.84&   $-$4.15&  2.70\\
$-$211.47, $-$211.65             & 12, 67      &      2.577& 0.005&    $-$0.251& 0.007&   $-$36.81&  1.84&   $-$6.02&  2.91\\
$-$211.95, $-$211.86             & 12, 67      &      2.547& 0.005&    $-$0.245& 0.009&   $-$43.10&  1.84&   $-$1.98&  4.28\\
$-$211.95, $-$212.07             & 12, 67      &      2.552& 0.005&    $-$0.243& 0.010&   $-$41.57&  1.85&   $-$2.92&  4.58\\
$-$212.43, $-$212.28             & 12, 6       &      2.546& 0.008&    $-$0.235& 0.023&   $-$29.41&  3.50&  $-$17.42& 12.27\\
 Weighted average                &             &      2.560& 0.002&    $-$0.257& 0.002&   $-$30.99&  0.48&      3.71&  0.75\\
                                 &             &           &      &            &      &           &      &          &      \\
 Feature 9                       &             &           &      &            &      &           &      &          &      \\
$-$211.47, $-$211.65             & 23, 67      &   $-$0.720& 0.006&    $-$2.522& 0.019&   $-$15.73&  2.87&      1.40&  9.95\\
$-$211.95, $-$211.86, $-$212.01  & 123, 67, 8  &   $-$0.693& 0.005&    $-$2.517& 0.014&    $-$8.64&  1.72&  $-$15.68&  4.57\\
$-$211.95, $-$212.07, $-$212.01  & 123, 67, 8  &   $-$0.690& 0.005&    $-$2.510& 0.014&    $-$7.86&  1.69&  $-$14.19&  4.49\\
$-$212.43, $-$212.28             & 12, 67      &   $-$0.684& 0.007&    $-$2.475& 0.022&    $-$5.71&  2.78&  $-$24.40& 10.65\\
$-$212.43, $-$212.49             & 12, 67      &   $-$0.687& 0.008&    $-$2.458& 0.027&    $-$7.36&  3.07&  $-$17.10& 11.95\\
 Weighted average                &             &   $-$0.699& 0.003&    $-$2.508& 0.010&    $-$9.44&  1.31&  $-$13.19&  3.87\\
                                 &             &           &      &            &      &           &      &          &      \\
 Feature 10                      &             &           &      &            &      &           &      &          &      \\
$-$212.91, $-$213.12             & 35, 6       &      1.984& 0.008&       0.709& 0.038&   $-$20.50&  9.41&     60.47& 34.96\\
$-$213.39, $-$213.33             & 3, 67       &      1.902& 0.008&       0.714& 0.021&    $-$5.33&  4.74&   $-$6.01& 13.80\\
$-$213.87, $-$213.75             & 3, 67       &      1.883& 0.008&       0.709& 0.020&    $-$4.05&  4.86&     11.92& 13.40\\
$-$213.87, $-$213.96             & 3, 67       &      1.882& 0.008&       0.701& 0.022&    $-$7.48&  6.05&   $-$1.40& 18.18\\
 Weighted average                &             &      1.923& 0.005&       0.710& 0.014&    $-$7.25&  3.32&      4.73&  9.82\\
                                 &             &           &      &            &      &           &      &          &      \\
 Feature 11                      &             &           &      &            &      &           &      &          &      \\
$-$211.95, $-$211.86             & 234, 67     &   $-$0.280& 0.004&    $-$1.320& 0.013&   $-$11.51&  2.53&  $-$20.28&  7.76\\
$-$211.95, $-$212.07             & 234, 67     &   $-$0.275& 0.004&    $-$1.316& 0.012&    $-$7.30&  2.53&  $-$19.95&  6.93\\
$-$212.43, $-$212.28             & 234, 67     &   $-$0.273& 0.004&    $-$1.335& 0.009&    $-$5.94&  2.53&  $-$19.54&  5.51\\
$-$212.43, $-$212.49             & 234, 67     &   $-$0.270& 0.005&    $-$1.336& 0.009&    $-$3.44&  2.54&  $-$20.72&  5.63\\
$-$212.91, $-$212.70, $-$212.85  & 12345, 67, 8&   $-$0.264& 0.004&    $-$1.327& 0.008&    $-$6.10&  1.80&  $-$20.18&  4.27\\
$-$212.91                        & 1234567     &   $-$0.270& 0.004&    $-$1.325& 0.008&   $-$12.01&  2.12&  $-$26.61&  5.21\\
$-$212.91, $-$213.12             & 12345, 67   &   $-$0.261& 0.004&    $-$1.323& 0.008&    $-$6.65&  2.12&  $-$25.41&  5.22\\
$-$213.39, $-$213.33             & 12345, 67   &   $-$0.271& 0.004&    $-$1.302& 0.007&    $-$8.97&  2.14&  $-$25.58&  5.26\\
$-$213.39, $-$213.54             & 12345, 67   &   $-$0.275& 0.004&    $-$1.305& 0.007&   $-$11.52&  2.14&  $-$27.97&  4.78\\
$-$213.87, $-$213.75, $-$213.70  & 12345, 67, 8&   $-$0.275& 0.004&    $-$1.274& 0.006&    $-$4.00&  1.71&   $-$3.96&  3.72\\
$-$213.87, $-$213.96, $-$213.70  & 12345, 67, 8&   $-$0.280& 0.004&    $-$1.271& 0.006&    $-$5.57&  1.71&   $-$1.67&  3.51\\
$-$214.36, $-$214.18, $-$214.54  & 2345, 6, 8  &   $-$0.286& 0.004&    $-$1.233& 0.007&   $-$10.66&  2.28&  $-$11.90&  4.34\\
$-$214.84                        & 245         &   $-$0.291& 0.007&    $-$1.200& 0.014&    $-$8.61&  4.18&   $-$4.47& 11.99\\
 Weighted average                &             &   $-$0.276& 0.002&    $-$1.285& 0.003&    $-$7.74&  0.85&  $-$14.71&  1.92\\
                                 &             &           &      &            &      &           &      &          &      \\
 Feature 12                      &             &           &      &            &      &           &      &          &      \\
$-$214.84, $-$214.60             & 13, 67      &   $-$0.327& 0.006&    $-$1.059& 0.016&   $-$24.88&  2.85&     12.85&  8.05\\
$-$214.84, $-$214.81             & 13, 67      &   $-$0.330& 0.006&    $-$1.030& 0.016&   $-$25.37&  2.85&     13.02&  8.05\\
$-$214.84, $-$215.02             & 13, 67      &   $-$0.328& 0.006&    $-$1.037& 0.016&   $-$24.55&  2.85&     12.05&  8.08\\
$-$215.32, $-$215.23, $-$215.38  & 1345, 67, 8 &   $-$0.341& 0.005&    $-$1.041& 0.015&   $-$14.25&  1.83&     16.99&  4.78\\
$-$215.32, $-$215.44, $-$215.38  & 1345, 67, 8 &   $-$0.330& 0.005&    $-$1.025& 0.017&   $-$11.78&  1.86&     19.12&  4.81\\
$-$215.80, $-$215.65             & 12345, 67   &   $-$0.348& 0.004&    $-$0.913& 0.008&    $-$8.40&  2.41&  $-$35.56&  8.67\\
$-$215.80, $-$215.86             & 12345, 67   &   $-$0.347& 0.004&    $-$0.910& 0.008&    $-$7.99&  2.28&  $-$31.00&  8.01\\
$-$216.28, $-$216.07             & 12345, 67   &   $-$0.350& 0.004&    $-$0.925& 0.007&    $-$7.53&  2.29&  $-$29.73&  6.28\\
$-$216.28                        & 1234567     &   $-$0.351& 0.004&    $-$0.921& 0.007&    $-$8.51&  2.29&  $-$21.35&  5.17\\
$-$216.28, $-$216.49             & 12345, 67   &   $-$0.354& 0.004&    $-$0.917& 0.007&   $-$10.47&  2.29&  $-$16.78&  5.06\\
$-$216.76, $-$216.70             & 1234, 67    &   $-$0.338& 0.005&    $-$0.930& 0.007&   $-$14.27&  2.51&   $-$4.57&  5.54\\
$-$216.76, $-$216.91             & 1234, 67    &   $-$0.332& 0.005&    $-$0.932& 0.007&   $-$10.92&  2.61&   $-$8.59&  6.74\\
$-$217.24, $-$217.12             & 1234, 67    &   $-$0.315& 0.005&    $-$0.977& 0.008&   $-$15.86&  2.84&   $-$7.36&  8.29\\
$-$217.24, $-$217.33             & 1234, 7     &   $-$0.312& 0.005&    $-$0.975& 0.008&   $-$13.11&  2.95&   $-$0.98& 10.47\\
$-$217.72, $-$217.54, $-$217.91  & 1234, 7, 8  &   $-$0.299& 0.005&    $-$1.007& 0.009&   $-$11.83&  2.05&     35.08&  4.96\\
$-$217.72, $-$217.75, $-$217.91  & 1234, 7, 8  &   $-$0.301& 0.006&    $-$1.006& 0.009&   $-$12.58&  2.13&     35.88&  5.13\\
$-$217.72, $-$217.96, $-$217.91  & 1234, 7, 8  &   $-$0.291& 0.006&    $-$1.010& 0.009&    $-$9.13&  2.17&     33.46&  5.18\\
$-$218.20, $-$217.96             & 34, 7       &   $-$0.278& 0.007&    $-$1.034& 0.014&   $-$22.59&  6.43&      3.98& 21.88\\
 Weighted average                &             &   $-$0.330& 0.002&    $-$0.957& 0.003&   $-$12.83&  0.87&      4.77&  2.31\\
                                 &             &           &      &            &      &           &      &          &      \\
 Feature 13                      &             &           &      &            &      &           &      &          &      \\
$-$216.76, $-$216.70             & 1234, 6     &     26.236& 0.006&   $-$15.531& 0.025&      10.02&  3.55&  $-$12.61& 13.60\\
$-$216.76, $-$216.91             & 1234, 6     &     26.226& 0.006&   $-$15.534& 0.025&       4.46&  3.55&  $-$14.03& 13.60\\
$-$217.24, $-$217.12             & 1234, 6     &     26.222& 0.006&   $-$15.523& 0.021&       4.56&  2.88&  $-$18.84& 10.30\\
$-$217.24, $-$217.33             & 1234, 6     &     26.229& 0.006&   $-$15.541& 0.019&       7.50&  2.70&  $-$27.22&  9.59\\
$-$217.72, $-$217.54             & 1245, 6     &     26.226& 0.005&   $-$15.527& 0.016&       1.24&  2.34&  $-$32.19&  7.41\\
$-$217.72, $-$217.75             & 1245, 6     &     26.235& 0.005&   $-$15.528& 0.016&       4.85&  2.34&  $-$32.35&  7.42\\
$-$217.72, $-$217.96             & 1245, 6     &     26.246& 0.005&   $-$15.531& 0.019&       9.50&  2.56&  $-$33.53&  8.45\\
$-$218.20, $-$217.96             & 125, 6      &     26.230& 0.008&   $-$15.521& 0.025&       5.54&  2.84&  $-$26.62&  9.60\\
$-$218.20, $-$218.17             & 125, 6      &     26.225& 0.006&   $-$15.505& 0.017&       3.94&  2.44&  $-$20.75&  6.89\\
$-$218.20, $-$218.38             & 125, 6      &     26.227& 0.006&   $-$15.510& 0.013&       4.76&  2.44&  $-$22.49&  5.45\\
$-$218.69, $-$218.60             & 1234, 6     &     26.249& 0.005&   $-$15.517& 0.009&       6.19&  2.71&  $-$19.41&  4.65\\
$-$218.69, $-$218.81             & 1234, 6     &     26.252& 0.005&   $-$15.526& 0.009&       7.58&  2.71&  $-$22.94&  4.56\\
 Weighted average                &             &     26.235& 0.003&   $-$15.523& 0.007&       5.78&  1.22&  $-$23.19&  3.14\\
                                 &             &           &      &            &      &           &      &          &      \\
 Feature 14                      &             &           &      &            &      &           &      &          &      \\
$-$218.20, $-$218.38             & 245, 67     &   $-$0.005& 0.006&    $-$2.722& 0.026&    $-$0.34&  4.01&  $-$13.81& 14.35\\
$-$218.69, $-$218.60             & 2345, 67    &      0.008& 0.005&    $-$2.742& 0.018&    $-$7.82&  3.31&   $-$8.35& 10.78\\
$-$218.69, $-$218.81             & 2345, 67    &      0.007& 0.005&    $-$2.745& 0.017&    $-$8.87&  3.20&  $-$10.84& 10.21\\
 Weighted average                &             &      0.002& 0.004&    $-$2.737& 0.015&    $-$5.18&  2.53&  $-$11.11&  8.47\\
                                 &             &           &      &            &      &           &      &          &      \\
 Feature 15                      &             &           &      &            &      &           &      &          &      \\
$-$219.65                        & 136         &   $-$0.447& 0.008&    $-$0.656& 0.027&   $-$27.69&  3.71&   $-$3.69& 14.80\\
$-$219.65, $-$219.44             & 13, 67      &   $-$0.452& 0.006&    $-$0.596& 0.024&   $-$29.55&  3.15&     28.29& 12.85\\
$-$219.65, $-$219.86             & 13, 6       &   $-$0.448& 0.008&    $-$0.655& 0.029&   $-$27.94&  4.02&   $-$3.36& 16.17\\
 Weighted average                &             &   $-$0.450& 0.007&    $-$0.631& 0.026&   $-$28.55&  3.57&      9.80& 14.41\\
                                 &             &           &      &            &      &           &      &          &      \\
 Feature 16                      &             &           &      &            &      &           &      &          &      \\
$-$219.65                        & 2346        &     26.558& 0.006&   $-$15.322& 0.018&       3.15&  6.11&  $-$16.78& 19.79\\
$-$219.65, $-$219.86             & 234, 6      &     26.557& 0.007&   $-$15.347& 0.020&       1.79&  6.88&  $-$53.06& 23.05\\
$-$220.13, $-$220.28             & 34, 6       &     26.537& 0.007&   $-$15.314& 0.016&   $-$29.65& 10.93&      2.50& 29.03\\
 Weighted average                &             &     26.548& 0.005&   $-$15.322& 0.012&    $-$5.79&  5.56&  $-$20.09& 17.14\\
                                 &             &           &      &            &      &           &      &          &      \\
\phantom{ IC 133}IC\,133 West:   &             &           &      &            &      &           &      &          &      \\
                                 &             &           &      &            &      &           &      &          &      \\
 Feature 1                       &             &           &      &            &      &           &      &          &      \\
$-$222.05                        & 123         & $-$283.137& 0.016&   $-$39.804& 0.068&    $-$8.90&  6.41&  $-$15.01& 25.72\\
$-$222.54                        & 123         & $-$283.155& 0.012&   $-$39.813& 0.029&   $-$15.72&  5.31&   $-$8.34& 13.94\\
 Weighted average                &             & $-$283.149& 0.010&   $-$39.812& 0.027&   $-$12.94&  4.09&   $-$9.85& 12.26\\
                                 &             &           &      &            &      &           &      &          &      \\
 Feature 2                       &             &           &      &            &      &           &      &          &      \\
$-$230.24, $-$230.04             & 23, 7       & $-$283.196& 0.010&   $-$44.740& 0.039&   $-$49.21&  5.10&     43.86& 18.24\\
$-$230.24, $-$230.25             & 23, 7       & $-$283.136& 0.011&   $-$44.799& 0.042&   $-$13.08&  5.71&     13.06& 20.36\\
$-$230.73, $-$230.88             & 23, 67      & $-$283.133& 0.006&   $-$44.763& 0.019&   $-$10.39&  3.05&     15.17&  8.88\\
$-$231.21, $-$231.09             & 23, 67      & $-$283.146& 0.005&   $-$44.786& 0.011&   $-$25.12&  2.53&     20.55&  5.68\\
$-$231.21, $-$231.30             & 23, 67      & $-$283.143& 0.005&   $-$44.754& 0.010&   $-$23.02&  2.53&     34.31&  5.11\\
$-$231.69, $-$231.51             & 23, 67      & $-$283.148& 0.005&   $-$44.742& 0.014&   $-$26.27&  2.56&     36.23&  7.61\\
$-$231.69, $-$231.72             & 23, 67      & $-$283.147& 0.006&   $-$44.744& 0.016&   $-$25.59&  2.98&     38.23&  8.83\\
 Weighted average                &             & $-$283.145& 0.003&   $-$44.761& 0.008&   $-$21.98&  1.52&     27.77&  3.92\\
                                 &             &           &      &            &      &           &      &          &      \\
 Feature 3                       &             &           &      &            &      &           &      &          &      \\
$-$233.13, $-$232.98             & 245, 67     & $-$283.599& 0.007&   $-$44.542& 0.016&   $-$48.41&  4.71&     51.07& 14.45\\
$-$233.13, $-$233.19             & 245, 67     & $-$283.600& 0.007&   $-$44.537& 0.016&   $-$46.06&  4.68&     52.18& 13.78\\
$-$233.61, $-$233.40             & 245, 67     & $-$283.597& 0.006&   $-$44.526& 0.023&   $-$49.00&  5.03&     63.57& 16.95\\
$-$233.61                        & 2456        & $-$283.594& 0.006&   $-$44.526& 0.023&   $-$40.14&  5.65&     62.39& 18.87\\
$-$233.61, $-$233.82             & 245, 6      & $-$283.592& 0.006&   $-$44.524& 0.023&   $-$35.68&  6.50&     65.99& 22.19\\
 Weighted average                &             & $-$283.597& 0.005&   $-$44.535& 0.013&   $-$45.38&  3.61&     55.96& 11.32\\
                                 &             &           &      &            &      &           &      &          &      \\
 Feature 4                       &             &           &      &            &      &           &      &          &      \\
$-$235.06, $-$234.88             & 4, 67       & $-$284.191& 0.010&   $-$38.836& 0.032&   $-$42.40&  5.91&     39.63& 18.96\\
$-$235.06, $-$235.09             & 4, 67       & $-$284.191& 0.010&   $-$38.814& 0.032&   $-$42.19&  5.86&     12.64& 18.18\\
$-$235.54, $-$235.30             & 4, 67       & $-$284.176& 0.010&   $-$38.828& 0.039&   $-$56.78&  6.40&     88.74& 24.04\\
 Weighted average                &             & $-$284.184& 0.007&   $-$38.826& 0.025&   $-$48.93&  4.33&     49.16& 14.69\\
                                 &             &           &      &            &      &           &      &          &      \\
 Feature 5                       &             &           &      &            &      &           &      &          &      \\
$-$236.50, $-$236.56             & 34, 6       & $-$284.134& 0.007&   $-$38.738& 0.022&   $-$41.98&  8.78&  $-$60.70& 27.66\\
$-$236.98, $-$236.77             & 34, 6       & $-$284.113& 0.006&   $-$38.815& 0.021&   $-$32.68&  6.60&  $-$83.46& 20.31\\
$-$236.98                        & 346         & $-$284.111& 0.006&   $-$38.814& 0.021&   $-$27.53&  6.60&  $-$79.32& 20.31\\
$-$236.98, $-$237.19             & 34, 6       & $-$284.117& 0.006&   $-$38.801& 0.022&   $-$46.71&  7.46&  $-$44.30& 22.94\\
$-$237.94, $-$238.03             & 34, 6       & $-$284.162& 0.006&   $-$38.754& 0.019&   $-$77.21&  7.46&     86.42& 22.35\\
$-$238.43, $-$238.25             & 34, 6       & $-$284.159& 0.007&   $-$38.837& 0.024&   $-$56.80&  9.34&     80.95& 31.10\\
$-$238.43, $-$238.46             & 34, 6       & $-$284.158& 0.007&   $-$38.836& 0.024&   $-$54.52&  7.07&     90.25& 23.38\\
$-$238.43, $-$238.67             & 34, 6       & $-$284.161& 0.007&   $-$38.836& 0.024&   $-$66.51&  7.97&     94.15& 26.41\\
 Weighted average                &             & $-$284.142& 0.003&   $-$38.781& 0.011&   $-$52.99&  3.84&      8.87& 11.96\\
                                 &             &           &      &            &      &           &      &          &      \\
 Feature 6                       &             &           &      &            &      &           &      &          &      \\
$-$243.24, $-$243.30             & 234, 6      & $-$284.125& 0.007&   $-$38.663& 0.027&   $-$16.59&  4.51&     13.99&  15.25\\
$-$243.72, $-$243.51             & 234, 6      & $-$284.133& 0.006&   $-$38.648& 0.022&   $-$20.41&  4.15&      6.09&  13.95\\
$-$243.72                        & 2346        & $-$284.141& 0.006&   $-$38.605& 0.020&   $-$26.38&  3.85&     35.54&  12.88\\
$-$243.72, $-$243.93             & 234, 67     & $-$284.146& 0.005&   $-$38.631& 0.020&   $-$31.47&  3.61&     18.40&  12.25\\
$-$244.20, $-$244.14             & 234, 67     & $-$284.133& 0.005&   $-$38.660& 0.018&   $-$32.06&  3.56&     45.64&  12.20\\
$-$244.20, $-$244.35             & 234, 67     & $-$284.138& 0.005&   $-$38.673& 0.018&   $-$36.55&  3.56&     29.62&  12.20\\
$-$244.68, $-$244.56             & 2345, 67    & $-$284.130& 0.005&   $-$38.657& 0.009&   $-$32.85&  3.67&     27.75&  12.23\\
$-$244.68, $-$244.77             & 2345, 67    & $-$284.134& 0.005&   $-$38.654& 0.009&   $-$37.01&  3.72&     42.90&  12.18\\
$-$245.16, $-$244.98             & 345, 7      & $-$284.114& 0.006&   $-$38.667& 0.010&   $-$37.46&  5.63&  $-$28.25&  19.44\\
$-$245.16, $-$245.19             & 345, 6      & $-$284.118& 0.006&   $-$38.664& 0.010&   $-$56.75&  7.36&     56.59&  23.24\\
$-$245.16, $-$245.40             & 345, 6      & $-$284.115& 0.006&   $-$38.666& 0.010&   $-$32.37&  8.43&     10.94&  27.81\\
$-$245.64, $-$245.40             & 345, 6      & $-$284.117& 0.006&   $-$38.657& 0.012&   $-$39.21&  8.55&     10.63&  28.63\\
$-$245.64, $-$245.61             & 345, 6      & $-$284.116& 0.006&   $-$38.652& 0.012&   $-$33.97& 10.21&    111.42&  35.90\\
 Weighted average                &             & $-$284.129& 0.002&   $-$38.657& 0.005&   $-$30.48&  1.82&     27.72&   6.13


\enddata

\tablenotetext{a} {Each row corresponds to a single spectral channel.  Since 
there are three different spectral resolutions (see Table 3), up to three 
different entries may appear in this column.  Weighted averages account for 
the fact that the channels listed are not always statistically independent.  
See text for details.}

\tablenotetext{b} {Epochs at which a given V$_{LSR}$ value appears.  See Table 
2 for the dates of observation.}

\tablenotetext{c} {East-west ($\Delta x$) and north-south ($\Delta y$) offsets 
from the reference feature at epoch 4.}

\tablenotetext{d} {East-west ($\Delta v_x$) and north-south ($\Delta v_y$) 
motions with respect to the reference feature.}

\end{deluxetable}


\clearpage

\begin{deluxetable}{crrrrrrrr}
\tabletypesize{\scriptsize}
\tablewidth{0pc}
\tablenum{6}
\tablecolumns{9}
\tablecaption{Channel versus Feature Proper Motions in IC\,133 Main}

\tablehead{
\colhead{Feature}&  
\multicolumn{4}{c}{Channel Proper Motions}& \multicolumn{4}{c}{Feature Proper Motions}\\
&  
\multicolumn{4}{c}{$\hrulefill$}& \multicolumn{4}{c}{$\hrulefill$}\\
& 
\colhead{$\Delta v_x$}& \colhead{$\sigma v_x$}& 
\colhead{$\Delta v_y$}& \colhead{$\sigma v_y$}&
\colhead{$\Delta v_x$}& \colhead{$\sigma v_x$}& 
\colhead{$\Delta v_y$}& \colhead{$\sigma v_y$}\\
%
& 
\multicolumn{4}{c}{($\mu$as yr$^{-1}$)}& \multicolumn{4}{c}{($\mu$as yr$^{-1}$)}
}

\startdata
\phantom{1}1&  $-$0.78& 1.99&   10.82& 6.34&     1.38& 3.11&    10.21& 5.58\\
\phantom{1}2&  $-$1.67& 1.44&    2.60& 2.59&  $-$2.08& 2.52&     1.03& 2.96\\
\phantom{1}3& $-$13.93& 1.47&   17.22& 3.39& $-$15.72& 1.57&    20.32& 2.26\\
\phantom{1}4&     6.42& 1.92&$-$10.36& 4.68&     0.48& 1.29&  $-$0.75& 2.65\\
\phantom{1}5& $-$16.98& 2.57&    9.28& 9.01& $-$17.05& 2.59&     6.77& 8.21\\
\phantom{1}6&     2.45& 1.63&   14.41& 5.28&     0.45& 3.67&    19.48& 7.65\\
\phantom{1}7& $-$25.79& 3.83&   34.05&15.21& $-$17.26& 2.62&    19.77& 9.12\\
\phantom{1}8& $-$30.99& 0.48&    3.71& 0.75& $-$31.33& 1.81&     6.93& 1.84\\
\phantom{1}9&  $-$9.44& 1.31&$-$13.19& 3.87&  $-$2.28& 1.95&  $-$5.78& 4.42\\
          10&  $-$7.25& 3.32&    4.73& 9.82& $-$23.34& 9.00&     0.98&15.57\\
          11&  $-$7.74& 0.85&$-$14.71& 1.92&  $-$7.20& 2.53& $-$12.97& 4.13\\
          12& $-$12.83& 0.87&    4.77& 2.31&      $-$\phantom{3}&  $-$\phantom{3}&      $-$\phantom{3}&  $-$\phantom{3}\\
          13&     5.78& 1.22&$-$23.19& 3.14&    10.39& 4.24& $-$23.42& 6.02\\
          14&  $-$5.18& 2.53&$-$11.11& 8.47&  $-$3.31& 5.47& $-$12.87&14.67\\
          15& $-$28.55& 3.57&    9.80&14.41& $-$27.07& 2.59&    15.20& 9.28\\
          16&  $-$5.79& 5.56&$-$20.09&17.14&     3.46& 3.43& $-$33.17& 8.43\\

\tablecomments{Computation of proper motions by averaging the motions in individual channels should be more robust against 
the effects of blending than computation of proper motions by fitting to mean feature positions at each epoch.  When fitting 
``features'': If the position offset for any epoch deviated by more than 3-$\sigma$ from the least squares fit to the data for 
all epochs, we doubled all position offset errors for that feature.  If deviations of more than 3-$\sigma$ remained, we rejected 
the feature as being too spectrally blended to fit.  There were seven features that required we double the errors and one feature 
that was rejected completely.  It should be emphasized that the results from ``feature'' fits are for comparison purposes only.}

\enddata


\end{deluxetable}


\clearpage

\begin{deluxetable}{llrcccc}
\tabletypesize{\scriptsize}
\tablewidth{0pc}
\tablenum{7}
\tablecolumns{7}
\tablecaption{OVRO MM Array Observations}

\tablehead{
\colhead{Date} &
\colhead{Transition} &
\colhead{Frequency} &
\colhead{Array\tablenotemark{a}} &
\colhead{Major\tablenotemark{b}} &
\colhead{Minor\tablenotemark{b}} &
\colhead{P.A.\tablenotemark{b}}\\
&
&
\colhead{(GHz)}&
\colhead{Configuration}&
\colhead{(arcsec)}&
\colhead{(arcsec)}&
\colhead{($^{\circ}$)}
}

\startdata

1998 April 28; May 24, 28      & CO 1$\to$0    & \phantom{,}115\phantom{GHz}& L & 5.8 & 4.0 &           $-$20.7\\
1998 December 18               & CO 1$\to$0    & \phantom{,}115\phantom{GHz}& H & 2.5 & 2.0 & \phantom{$-$}16.4\\
1999 January 6, 7              & HCN 1$\to$0   & 89\,\phantom{GHz}          & H & 3.3 & 2.3 &           $-$89.2\\
                               & HCO$+$ 1$\to$0& 89\,\phantom{GHz}          & H & 3.3 & 2.0 & \phantom{$-$}89.9\\
                               & CO 2$\to$1    & \phantom{,}231\phantom{GHz}& H & 1.0 & 0.8 & \phantom{$-$}73.0\\
1999 February 2, 3             & CO 1$\to$0    & \phantom{,}115\phantom{GHz}& UH& 1.1 & 0.8 & \phantom{$-$}75.5


\enddata

\tablenotetext{a} {Resolutions: L is low, H is high, and UH is ultrahigh.}

\tablenotetext{b} {Restoring beam axes and position angle.  Position angles 
are measured east of north.}

\end{deluxetable}


\clearpage

\begin{deluxetable}{lccccc}
\tabletypesize{\scriptsize}
\tablewidth{0pc}
\tablenum{8}
\tablecolumns{6}
\tablecaption{Comparison with Galactic Water Vapor Maser Sources} 

\tablehead{
\colhead{Source} &
\colhead{V$_{LSR}$ spread} &
\colhead{Peak F$_{\nu}$} &
\colhead{Distance} &
\colhead{Peak F$_{\nu}$ at 850 kpc\tablenotemark{a}} &
\colhead{Expansion Velocity}\\
&
\colhead{(km s$^{-1}$)}&
\colhead{(Jy)}&
\colhead{(kpc)}&
\colhead{(Jy)}&
\colhead{(km s$^{-1}$)}
}

\startdata
W49$\,$N\tablenotemark{b}  & $\approx$ 500& 10000 (June 1978)                    & 11.4$\pm$1.2                & 1.8\phantom{0}& 55\\
W51$\,$M\tablenotemark{c}  & $\approx$ 100& \phantom{1}2000 (Nov. 1977)          & \phantom {1}7.0$\pm$1.5     & 0.14          & 40\\
W51$\,$N\tablenotemark{d}  & $\approx$ 100& \phantom{1}1000 (Dec. 1977,\phantom{,} & \phantom{1}8.3$\pm$2.5    & 0.10          & 43\\
                           &              & \phantom{10000 (}Nov. 1979)           &                             &\\
SgrB2$\,$N\tablenotemark{e}& $\approx$ 100& \phantom{10}400 (Dec. 1980$\,$)       & \phantom{1}7.1$\pm$1.5     & 0.03          & 45


\enddata

\tablenotetext{a} {This assumes a distance of 850 kpc to IC\,133, which is the 
average of the 13 values in Table 1.}

\tablenotetext{b} {See Walker, Matsakis, \& Garcia--Barreto (1982) and 
Gwinn, Moran, \& Reid (1992).  The authors fit to outflow from a single source 
and found an expansion velocity that is approximately constant out to a 
distance of $1''$ from the expansion center and increases rapidly after 
that.  We adopted an average expansion velocity for the 105 maser features.}

\tablenotetext{c} {See Genzel et al. (1981b).  Maser motions appeared to be 
random and isotropic in this source, so the authors employed the method of 
statistical parallax to determine the distance.  We adopted an expansion 
velocity equal to one half the radial velocity spread of features with proper 
motions.}

\tablenotetext{d} {See Schneps et al. (1981) and footnote c.}

\tablenotetext{e} {See Reid et al. (1988).  The authors fit to outflow 
from a single source and found an expansion velocity of 45$\pm$9 km s$^{-1}$.}

\end{deluxetable}


\clearpage


\end{document}